\def\,{\thinspace}

\def\kms{km\thinspace s$^{-1}$}
\def\Lsun{L$_\odot$}
\def\Msun{M$_\odot$}

\font\sc=cmr10

\def\Htwo{{\hbox {H$_2$}}}

\def\Lco{{\hbox {$L_{\rm CO}$}}}

\def\Lfir{{\hbox {$L_{\rm FIR}$}}}
\def\Ico{{\hbox {$I_{\rm CO}$}}}

\def\Tmb{{\hbox {$T_{\rm mb}$}}}
 
\def \deg{^\circ}

\def \sqr#1#2{{\vcenter{\hrule height.#2pt
	\hbox{\vrule width.#2pt height#1pt \kern#1pt
		\vrule width.#2pt}
	\hrule height.#2pt}}}

\documentstyle[12pt,aasms]{article}
\tighten
\def\Irat{\hbox{$I$(2-1)/$I$(1-0)}}
\def\Tb#1#2{\hbox{$T_b$(#1-#2)}}
\def\Trat{\Tb21/\Tb10}
\def\Lcs{{\hbox {$L_{\rm CS}$}}}
\def\LprimeCO{{\hbox {$L^\prime_{\rm CO}$}}}
\def\Lprimeco{{\hbox {$L^\prime_{\rm CO}$}}}
\def\Lcop{  { \hbox { $L^\prime_{\rm CO}$ }  }  }
\def\Lco {  { \hbox { $L_{\rm CO}$} }  }
\def \Mdyn{{\hbox {$M_{\rm dyn}$}}}
\def \dV{$\Delta V$}
\def \mhtwo{M(\Htwo)}
\def \MH2{M(\Htwo)}
\def \MiH2{M_i(\Htwo)}
\def \nH2{n(\Htwo)}
\def \nbarH2{\bar{n}}
\def \niH2{n_i(\Htwo)}
\def \fH2{f(\Htwo)}
\def \H2{\Htwo}
\def\Htwo{{\hbox {H$_2$}}}
\def\htwo{{\hbox {H$_2$}}}
\def \ts{\thinspace}
\def\,{\thinspace}
\def\Msun{M$_\odot$}
\def \msun {\hbox{M$_\odot$}}
\def \Lsun {\hbox{L$_\odot$}}
\def \lsun {\hbox{L$_\odot$}}
\def\Lfir{{\hbox {$L_{\rm FIR}$}}}
\def\lfir{{\hbox {$L_{\rm FIR}$}}}
\def \Tb{$T_b$}
\def \kms{{\hbox{km\,s$^{-1}$}}}
\def \Kkmspc{K\,\kms\,pc$^2$} 
\def \kkmspc{K\,\kms\,pc$^2$} 
\def \microns {$\mu$m}
\def \Microns {$\mu$m}
\def \micron {$\mu$m}
\def \Micron {$\mu$m}
\def\date{\number\day\space \ifcase\month\or
	January\or February\or March\or April\or May\or June\or 
	July\or August\or September\or October\or November\or December\fi
	\space\number \year}
\received{March 28, 1996}

\journalid{337}{15 January 1995}
\articleid{11}{14}
\slugcomment{}

\begin{document}
\title{ The Molecular Interstellar Medium in Ultraluminous Infrared Galaxies}
\author{P. M. Solomon}
\affil{Astronomy Program, State University of New York, Stony Brook, NY 11794}
\author{D. Downes}
\affil{Institut de Radio Astronomie Millim\'etrique, 
 38406 St.\ Martin d'H\`eres, France}
\author{S. J. E. Radford}
\affil{National Radio Astronomy Observatory, Tucson AZ 85721}\and
\author{J. W. Barrett}
\affil{Astronomy Program, State University of New York, Stony Brook, NY 11794}

\begin{abstract}
\setcounter{page}{1}
We present observations with the IRAM 30\,m telescope of CO in 
a large sample of ultraluminous IR galaxies out 
to redshift $z = 0.3$.  
 Most of the ultraluminous galaxies in this sample are interacting, but not 
completed mergers. 
 The CO(1--0) luminosity of all but one of the ultraluminous
 galaxies is high, with values of 
log (\Lprimeco /\Kkmspc ) $= 9.92\, \pm 0.12$ . The extremely small
dispersion of only 30 \% is less than that of the far infrared luminosity.
The integrated CO line intensity is strongly correlated with the
100\,$\mu$m flux density, as expected for a black body model in which the 
 mid and far IR radiation is optically thick. 
We use this model to derive sizes of the FIR and CO emitting regions and
the enclosed dynamical masses. 
 Both the IR and CO emission originate in regions a few hundred parsecs
in radius.  The median value of
$L_{\rm FIR}/\LprimeCO $ = 160~\Lsun/\Kkmspc , within
a factor of two or three of the black body limit for the observed far IR temperatures.  
 The entire ISM is a scaled up version of a normal galactic 
disk
with the ambient densities  a factor of 100 higher, making even the intercloud
medium  a molecular region.
 We compare three different techniques of \htwo\ mass estimation and
 conclude that the 
ratio of gas mass to CO luminosity is about a factor of  four times lower than
for Galactic molecular clouds, but that the gas mass is a
large fraction of the dynamical mass.
 Our analysis of CO emission from ultraluminous
galaxies reduces the \htwo\,mass from previous estimates of 2 --5 $\times 10^{10}\msun$ to
0.4--1.5 $\times 10^{10}\msun$, which is in the range found for molecular 
gas rich spiral galaxies. 
A collision involving a molecular gas rich spiral could lead to an ultraluminous galaxy
powered by  central starbursts triggered by  the compression of  infalling preexisting GMC's.

The extremely dense 
molecular gas in the center of an ultraluminous galaxy
 is an ideal stellar nursery for a huge starburst. 

\end{abstract}


\def\Irat{\hbox{$I$(2-1)/$I$(1-0)}}
\def\Tb#1#2{\hbox{$T_b$(#1-#2)}}
\def\Trat{\Tb21/\Tb10}
\def\Lcs{{\hbox {$L_{\rm CS}$}}}
\def\LprimeCO{{\hbox {$L^\prime_{\rm CO}$}}}
\def\Lcop{  { \hbox { $L^\prime_{\rm CO}$ }  }  }
\def\Lco {  { \hbox { $L_{\rm CO}$} }  }
\def \Mdyn{{\hbox {$M_{\rm dyn}$}}}
\def \dV{$\Delta V$}
\def \Htwo{H$_2$}
\def \htwo{H$_2$}
\def \msun{{\rm M}_\odot}
\def \MH2{M(\Htwo)}
\def \mhtwo{M(\Htwo)}
\def \MiH2{M_i(\Htwo)}
\def \nH2{n(\Htwo)}
\def \nbarH2{\bar{n}}
\def \niH2{n_i(\Htwo)}
\def \fH2{f(\Htwo)}
\def \H2{\Htwo}
\def \ts{\thinspace}
\def\,{\thinspace}
\def\Lsun{L$_\odot$}
\def \Tb{$T_b$}
\def \Kkmspc{K\,\kms\,pc$^2$} 
\def \kkmspc{K\,\kms\,pc$^2$} 

\font\sc=cmr10
\font\Bigbf=cmbx12 scaled \magstep2
\font\bigbf=cmbx12 scaled \magstep1
%
\parskip=0pt
\hfuzz=10pt
\section{INTRODUCTION: A NEW CO STUDY OF ULTRALUMINOUS GALAXIES} 
We report here an observational study of the molecular content of a 
large sample of extremely luminous
infrared galaxies. Ultraluminous galaxies, those with 
IR luminosities\footnote{
We use $H_0 = 75$\,\kms\,Mpc$^{-1}$ and $q_0 = 0.5$ throughout this paper.
}  
$>$   10$^{12}$\,\Lsun\, (cf.  Wright, Joseph and Meikle, 1984; Sanders et al.\ 1988a),
are the most luminous objects in the local universe.  They typically radiate
90\% or more of their energy in the far infrared.  
Most of these objects were discovered by the IRAS survey and many were
previously uncatalogued.  All ten of the brightest (nearest) 
ultraluminous galaxies (Sanders et al. 1988a) are either
merging systems or have tidal tails indicating a recent merger.   
Besides the very large IR luminosity, the molecular interstellar medium 
in such galaxies differs from that in normal spiral galaxies in  
several fundamental respects.  First, even though these galaxies have high 
CO luminosities and molecular masses, the ratio of far infrared 
to CO luminosity  is about an order of magnitude
 greater than for normal spiral galaxies (e.g., Sanders et al.\  1986;
Solomon \& Sage 1988). In star formation models, this implies a much 
higher star formation rate per Solar mass of molecular gas than in normal 
galaxies, even gas rich spirals.  Second, a
large fraction of their molecular gas is at densities much higher than in 
ordinary giant molecular clouds, as shown by HCN observations which trace gas 
at densities $> \, 10^5$\,cm$^{-3}$  (Solomon, Downes, \& Radford 1992a).      
Third, direct interferometric measurements of nearby ultraluminous galaxies 
(e.g., Scoville et al.\ 1991; Radford et al.\ 1991b) 
show most of the CO, HCN, and molecular mass is
concentrated in a small central region (less than 1\,kpc).  Indeed, estimates
of the molecular mass in the central condensation often equal the dynamical
mass (e.g., Scoville et al.\ 1991).  This conflict between dynamical mass and 
H$_2$ mass derived from CO luminosity leads to a new interpretation of the 
CO luminosity (Downes, Solomon, \& Radford 1993; hereafter DSR) 
for a medium which may fill a 
disk or sphere in the central few hundred parsecs of a galaxy.  

 In this paper we present the results of a systematic CO(1--0) survey with 
the IRAM 30\,m telescope of a large sample of ultraluminous galaxies out to 
$z = 0.27$. This is part of a multiline study  with a goal of 
understanding the difference between the star-forming environment in 
ultraluminous galaxies and large spiral galaxies. 
The observational data are in section 2.  We derive CO and IR luminosities and
temperatures in section 3 and compare them with the values for normal 
spirals.  We have previously suggested that for the extreme conditions of  
ultraluminous galaxies, the ratio of IR to
CO luminosity approaches the ratio expected for a black body (DSR)
This implies the dust is optically thick even at far~IR wavelengths 
(see also Condon et al.\  1991) and the ratio is not a simple indicator of  
luminosity to molecular mass. In section 4, we review the black body model 
and discuss the validity of our black-body approximation for the central 
regions. In section 5, we derive the dynamical mass from the model and the 
observations, and compare it with the H$_2$ mass derived from CO luminosity. 
We also derive a lower limit to the molecular mass by assuming the CO  is optically thin, and we
compare the mass calculated by different  approximations. We discuss the limits of the model, the
extent to which gas may be a dominant part of the dynamical mass in the centers of 
these galaxies and the implications for the nature of the luminosity source.

\section{OBSERVATIONS}
\subsection{The Sample and Observing Methods}
\subsubsection{Source Selection}
The sample contains 37 infrared luminous galaxies in the redshift range z = 0.03 to 0.27. 
 Of these 11, including the well known sources 
Arp\,220 and Mrk\,231,
have 60\,$\mu$m fluxes 
$S_{60} > 5.0$\, Jy and are part of the  nearby bright galaxy sample
(Sanders et al.\ 1988a;  Sanders, Scoville, \& Soifer 1991).   They have previously been
observed in CO by  several groups.  
 Twenty galaxies were chosen from a redshift survey (Strauss 
et al.\ 1992) of all IRAS sources with 60\,$\mu$m fluxes 
$S_{60} > 1.9$\, Jy.  We selected  primarily the most far infrared
 luminous sources  (see equation 4) at
$\delta > -10\deg$;   a few
lower luminosity sources were also included. 
We excluded a few sources from the redshift
survey whose unrealistically high 100/60\,$\mu$m flux ratios indicate they
are probably contaminated by Galactic emission. 
In order to include more galaxies at very high infrared luminosity we also observed 
7 sources with 60\,$\mu$m fluxes between  1.2 and 1.9\,Jy from the extension of
the redshift survey.  ( Fisher et al., 1995) These are the most distant,  $0.15  \le \,z
\le \, 0.27$,  galaxies in the sample.
    
Although not complete, our sample includes a large fraction of the most
luminous IRAS galaxies in the northern sky. Of the ten most luminous galaxies in
the redshift survey above $\delta > -10\deg$, our sample includes eight.
One of the missing sources is the radio-loud quasar 3C273.
Twenty galaxies in our sample are more luminous than the prototype IR
galaxy Arp\, 220 and 
 25 have a far infrared
luminosity , $\lfir \,> 1 \times 10^{12}$ \lsun,  obtained from the color corrected  60
and 100\,$\mu$m flux (see equation 4). 
 The most luminous object in the sample, 14070+0525, with 
$4 \times 10^{12}$\,\Lsun\,  in the far~IR, is also the most distant.

In our sample, the mean far~IR color  
$S_{60}/S_{100} = 0.81$, is  slightly cooler than in the smaller sample 
studied by Sanders et al.\ (1988a), where the mean
$S_{60}/S_{100} = 0.93$ (Sanders, private communication).  
Only 19\% of the galaxies in our sample have  
$S_{60}/S_{100} > 1.0$, while 50\% have $S_{60}/S_{100} > 1.0$ in Sanders 
et al.'s  sample.  A possible explanation is that although  most of  our 
sample was taken from a redshift survey of 60\,$\mu$m sources, 
we selected candidates based on total far infrared luminosity  including the
contribution from 100 \Microns.
The sample covers higher redshifts than the earlier studies 
(Sanders et al.\ 1988a; 1991).  {\bf Figure~1} shows the redshift
distribution for the galaxies described here, which extends to 
$cz = 80000$\,\kms .  For comparison, Sanders, et al.\ (1991) included 
three ultraluminous galaxies in their highest
range of redshift,  $cz = 15000$ -- 25000\,\kms .  
In the redshift range  $c z=$ 15000 -- 80000\,\kms , our sample has 
25 galaxies with 60\,$\mu$m fluxes between 1.0 and 5\,Jy.  Of these, 
22 galaxies have a FIR luminosity $L_{\rm FIR} >  1 \times 10^{12}$\,\Lsun . 

\subsubsection {Positions of the IRAS Galaxies}
With the IRAM 30\,m telescope, which has a beamwidth of
22$''$ at 115 GHz and 13$''$ at 230 GHz, CO observations
require source positions accurate to a few arcsec.
This is more accurate than positions in the IRAS catalogs (Moshir
et al.\ 1992; Beichman et al.\ 1988), which typically have error
ellipses with  major axes of 15$''$ -- 20$''$.  For the faintest 
sources, the IRAS positions may be no better than 30$''$.
Hence we determined positions of candidate galaxies near
the center of the IRAS error ellipse for each source
from the Palomar Sky Survey prints with a measuring engine.
The weakest sources have visual magnitudes near the Sky Survey limit 
(mag 17 to 18).  We could usually identify the most likely galaxy by 
inspection and we detected CO in all  of these with no exceptions. 
  {\bf Table~1} gives the positions where we
detected CO. We estimate the  accuracy to be $\pm3''$ for sources with positions from the 
Sky Survey. For a few of the  stronger sources we used VLA continuum 
positions (Condon et al.\  1991). 

 Some  potential
sources  selected from the extension to the redshift survey with 60\,$\mu$m fluxes
between  1.2 and 1.9\,Jy are in confused fields where we found several very faint
candidates.   In 5 cases we stopped looking after one or two tries due to limited
observing time.  The limits in these cases were not low enough to be interesting given 
the very weak far IR flux.  In other words we simply did not invest the observing time
required to reach the expected integrated intensity given the low 100 $\mu$m flux
(see Fig. 4, section 4), and the confused position.
These sources are not included in the sample.

\subsubsection { CO Observations}
The CO observations were made with 
the IRAM 30\,m telescope on Pico Veleta near 
Granada, Spain.  All observations were done with a wobbling secondary with a
throw of 120$''$ -- 240$''$ in a double switching mode, alternating on--off
and off--on.  The SIS receivers had typical SSB noise temperatures
of 130 -- 230\,K .  The 512 $\times$ 1\,MHz filter banks 
covered 1400 -- 1600 \kms\ at the redshifted CO(1--0) line and the data 
were smoothed to resolutions of 8\,MHz ($\approx 24$\,\kms) 
or 16\,MHz ($\approx 48$\,\kms) for analysis.
To test the reality of the weakest lines, we used two local oscillator 
settings, shifting the lines by 200 -- 300\,\kms .  
The data were calibrated with cold and ambient loads and
pointing was checked on planets and quasars. Atmospheric opacities
were typically $<0.1$ at the observing frequencies of 91 -- 110\,GHz. At 
these frequencies, the telescope's forward efficiency is 90\% and its
main beam efficiency is 60\%.  In this paper, spectra and intensities, \Ico , 
are in main beam brightness temperature,  $T_{\rm mb}$, which is appropriate 
for small sources. For the 30\,m telescope, $T_{\rm mb} = $1\,K corresponds
to a flux density of 4.5\,Jy from a point source in the 3mm band.   

Spectra of  the sources are shown in {\bf Fig.~2} and integrated 
intensities are listed in {\bf Table~1}.  
A few of these spectra have been published previously 
(Radford, Solomon \& Downes 1991a; Solomon, Downes \& Radford 1992a).
The peak intensities  
range from $T_{\rm mb}= 2$\, mK for the most distant object at $z = 0.265$ 
to  $\sim 100$\,mK for the nearest sources.  The CO redshifts agree well with 
the optical redshifts; the difference between the two is typically
$\leq 30$\,\kms.  The linewidths in {\bf Table~1} are either the full width at 
half maximum of Gaussian fits or one-half the full
width to zero intensity  for line profiles that are strongly non-gaussian.
Unlike spectra from normal spirals, many of these profiles are centrally 
peaked. 

In addition to CO data, we present R-band CCD images of 12 of the galaxies
between RA 10h and 23h, taken at the University of Hawaii 2.2 m telescope.   
The galaxy morphology shown by these images 
is discussed in Section~6; for easy reference, we place
our CO spectra next to the images  in order of decreasing redshift ({\bf Fig.~7}).

\section {CO AND INFRARED LUMINOSITIES}
{\bf Table~2} lists the CO and far~IR luminosities of the sample galaxies. 
Here we describe how we calculate luminosities from observed fluxes and we 
compare the luminosities with those of normal galaxies. 
\subsection{CO Luminosities}
\subsubsection{Basic Expressions}
The CO line luminosity  can be expressed several ways. Monochromatic
luminosity $ L(\nu_{\rm rest})$, observed flux density 
$S(\nu_{\rm obs})$, and luminosity distance $D_L$ are related by 
$\nu_{\rm rest} L(\nu_{\rm rest}) = 4 \pi D_L^2\, \nu_{\rm obs}\,
S(\nu_{\rm obs})$, so 
\begin{equation}
\Lco = 1.04 \times 10^{-3} \, S_{\rm CO}\, \Delta V\
	\nu_{\rm rest} (1+z)^{-1}\, D_L^2 \ \ \ \ \ ,                
\end{equation}
where  $\Lco $ is the CO line luminosity in \Lsun, $S_{\rm CO}\, \Delta V $ 
is the velocity integrated flux in Jy\,\kms, 
$\nu_{\rm rest} = \nu_{\rm obs} (1+z)$ is the 
rest frequency in GHz, and $D_L$ is the luminosity distance in Mpc.

Often CO line luminosity is expressed as velocity integrated source brightness 
temperature, $T_b \, \Delta V$, times source area, 
$\Omega_{\rm s}  D_A^2$, where $\Omega_{\rm s}$ is the solid angle subtended 
by the source and $D_A = D_L/(1+z)^2$ is the angular size distance.  
The observed integrated line intensity, $\Ico = \int \Tmb\,dV$, is obtained 
from the beam-diluted brightness temperature.  This must be corrected for 
redshift to get the intrinsic source brightness temperature,
 $T_b \, \Delta V\, \Omega_{\rm s}  = \Ico \Omega_{\rm s \star b}
(1+z)$, where $\Omega_{\rm s \star b}$ is the solid angle of the source 
convolved with the telescope beam.  Then the line luminosity
$\Lcop = T_b \, \Delta V \, \Omega_{\rm s} D_A^2 
= \Omega_{\rm s \star b} D_L^2 \Ico (1+z)^{-3}$, or
\begin{equation}
 \Lcop = 23.5 \ \Omega_{\rm s \star b}\, D_L^2\, \Ico\, (1+z)^{-3} 
\end{equation}
when $\Lcop$ is in \Kkmspc, $\Omega_{\rm s \star b}$ is in arcsec$^2$,
$D_L$ is in Mpc, and $\Ico$ is in K\,\kms. 
If the source is much smaller than the
beam, $\Omega_{\rm s \star b} \approx \Omega_{\rm b}$. Here we see that for a 
fixed beam size and source luminosity,
the integrated line intensity does {\it not} scale as $D_L^{-2}$, but rather as
$(1+z)^3 D_L^{-2}$ (Solomon, Radford, \& Downes 1992). 
The line luminosity, $\Lcop$, can also be expressed for a source of any size
in terms of the total line flux, $\Lcop = (c^2/2k) S_{\rm CO}\, \Delta V \,
 \nu_{\rm obs}^{-2} \, D_L^2\, (1+z)^{-3}$, or 
\begin{equation}
\Lcop = 3.25 \times 10^7 \, S_{\rm CO}\, \Delta V \,
	\nu_{\rm obs}^{-2}\, D_L^2 \, (1+z)^{-3} \ \ \ .  
\end{equation} 
with $S_{\rm CO}$ $\Delta V$ in Jy \kms , $\nu_{\rm obs}$ in GHz,
and $D_L$ in Mpc. 

The quantity $L^\prime_{\rm CO}$ is useful because 
it is surface brightness times area, in 
brightness temperature units.  Thus two lines with the same $T_b$ and extent 
will have the same $L^\prime_{\rm CO}$, regardless of transition or line 
frequency.  Conversely, the $L^\prime_{\rm CO}$ ratio of two lines 
is the average over the source of the lines' intrinsic $T_b$ ratio, which 
is an indicator of physical conditions in the gas.  

\subsubsection {CO Luminosities and Gas Masses for this Sample}
{\bf Table~2} lists the CO luminosities and the nominal gas 
masses $M^\prime({\rm H}_2)$ computed with a standard Milky Way  
H$_2$ mass--to--CO luminosity ratio
of 4.6\,\Msun /\Kkmspc\ (Solomon et al.\  1987).
The average for the sample is  $\Lcop =8\times 10^{9}$ \Kkmspc, 
or $ 4 \times 10^{10}$\,\Msun\ of molecular gas.  These are  
$\sim 20$ times greater than the CO luminosity and molecular mass of the 
Milky Way interior to the Solar circle (Solomon \& Rivolo 1989). Previous  
CO surveys of spiral galaxies show there are many isolated, non-interacting 
spirals with CO luminosities 2 -- 5 times higher than the Milky Way. 
Examples include NGC\,6946, NGC\,7479, and NGC\,1530 (cf. Solomon \& Sage 1988).  
A particularly gas rich, isolated galaxy with  normal far~IR properties is 
NGC\,3147. Its CO luminosity and molecular mass are 15 times larger than the
Milky Way's and close to the average for these ultraluminous IR galaxies. High
molecular masses appears, however, 
to be more common in interacting galaxies than in normal galaxies.
In a group of 29 interacting galaxies with separations $< 5\ D_{25}$ 
and some morphological disturbances,  Solomon \& Sage (1988) found
15 had molecular masses greater than $5 \times 10^{9}$\,\Msun. 
Below (section~5) we derive the molecular mass of ultraluminous galaxies 
by several methods, including estimates of the dynamical mass and of
the \htwo\ mass for optically thin CO emission.  We conclude 
the standard  Milky Way $M({\rm H}_2$)/$L^\prime_{\rm CO}$ 
ratio overestimates the \htwo\ masses of these galaxies by a factor of three. 
\subsection {Infrared Luminosities and Dust Temperatures}
>From the fluxes in the IRAS Faint Source Catalog we calculated FIR luminosities 
({\bf Table~2}).
\begin{equation}
L_{\rm FIR} =
3.94\times 10^5\, (2.58\, S_{60} + S_{100})\ r(S_{60}/S_{100})\ D_L^2,
\end{equation}
\noindent
with  $L_{\rm FIR}$ in \Lsun , $S$ in Jy, and $D_L$ in Mpc.
The color correction, $r$, is a function of the 
60 -- 100\,\micron\ flux ratio and the assumed dust emissivity (Lonsdale 
et al.\ 1985).  This is a multiplicative factor between 1.5 and 2.1 that
allows the FIR luminosity to be derived from the 60 and 100\,\micron\ 
fluxes.

Note, we use far~IR luminosity, $L_{\rm FIR}$ rather than IR luminosity, 
$L_{\rm IR}$, which includes the 25 $\mu$m flux (Sanders et al.\ 1991). 
For the faintest galaxies in the sample the 25  $\mu$m flux has a large uncertainty
and in any case contributes only about 20\% to the total 
 IR luminosity except for four warm objects with
25/60
\Micron\ ratios $>\,0.2 $ where 
$L_{\rm IR}$ is about 50\% higher than $L_{\rm FIR}$.
In terms of IR luminosity  the number of  sources in the sample  with 
$L_{\rm IR} \,> 1 \times  \, 10^{12} $\lsun  remains at 25; there are another 7 with 
$L_{\rm IR} \,>  0.8 \times  \, 10^{12} $\lsun. 
 
The dust temperature derived from the far~IR colors depends
on the assumed emissivity for optically thin dust but is
independent of emissivity for an optically thick source. 
While the dust in normal galaxies, including the Milky Way, is 
transparent in the far~IR, 
there is strong evidence (next section and Condon et al.\ 1991) the dust 
in ultraluminous galaxies is opaque, even at 60 and 100\, $\mu$m. 
We used, therefore, a black body emissivity index,
$n = 0$, to calculate dust temperatures and color corrections. This gives 
higher dust temperatures than the optically thin approximation.
The observed dust temperatures were also multiplied by ($1+z$) to obtain 
rest frame temperatures.

{\bf Figure~3} is a diagram of FIR luminosity vs.~CO luminosity for the 
galaxies in our sample.  For comparison, 
some well-known galaxies and the trend for normal
and weakly interacting spirals (Solomon \& Sage 1988) are also shown.
Ultraluminous galaxies have a systematically
 higher infrared to CO luminosity ratio. In the following section 
we develop a model which explains the high ratio and shows that there
 is a clear upper limit to \Lfir/\Lco. The model also leads to a re-interpretation
of the molecular mass determination for ultraluminous galaxies.
\section{BLACK BODY MODEL}
We argued previously (DSR) that for ultraluminous galaxies, the tight 
correlation of CO line intensity and 100\,$\mu$m flux strongly indicates  
the dust is optically thick at 100\,$\mu$m. 
The essential steps in that argument were:

{\it a) Small sizes of CO regions imply black bodies in the FIR.}
Interferometers show the molecular gas in ultraluminous galaxies is in small 
central regions. Examples are Arp\,220, with a CO and HCN core radius of 
320\,pc (Scoville et al.\ 1991; Okumura et al.\  1991; Radford et al.\   
1991b); Mrk\,231, with a CO source radius $0.7$\,kpc 
(Bryant \& Scoville 1996); and 17208 -- 0014, with a CO 
radius $<1.2$\,kpc (Planesas, Mirabel, \& Sanders 
1991). The small CO sizes and the H$_2$ masses estimated from the CO 
luminosities yield column densities $n({\rm H}_2)\geq 10^{24}$\,cm$^{-2}$.  For
a Galactic gas-to-dust mass ratio of 100 -- 200, the standard relation
for far~IR dust opacity (e.g., Hildebrand 1983) then yields 
$\tau_{\rm dust}>1$ at 100\,$\mu$m, so the far~IR source is optically thick.

{\it b) Tight correlation of \Ico\ with 100\,$\mu$m flux implies black body 
radiation}. For the ultraluminous galaxies, CO integrated line 
intensity, \Ico , is 
tightly correlated with 100\,$\mu$m flux density ({\bf Fig.~4}).  The scatter 
in this correlation 
is about a factor of two.  As we showed earlier (DSR), 
this flux-flux correlation can be understood if both 
the CO and the dust are optically thick.  Not only does CO line intensity 
vary linearly with 100\,$\mu$m flux density, but the observed 
constant of proportionality agrees with  
the black body model.  To demonstrate this, the argument in Downes, Solomon, 
\& Radford (1993), that was given in terms of the monochromatic 
100\,$\mu$m luminosity, $L_{100}$, will now be reformulated in terms of the 
total far~IR luminosity $L_{\rm FIR}$. In our earlier presentation of this 
argument, we assumed the CO linewidth $\Delta V$ was the same on on all lines 
of sight; in this version, we allow for a velocity filling factor $f_V$.

Assume the CO traces a huge cloud of area filling factor 
unity and the dust is opaque for $\lambda \leq 100\,\mu$m. 
The source may be a disk, a torus, a bar, or have an irregular 
shape.  For illustration, we derive the luminosity ratio for a sphere, but as 
long as the CO and far~IR sources cover the same area, the FIR/CO luminosity 
ratio is independent of source shape.  For an optically thick sphere 
of temperature $T_d$ at its outer radius $R$,  
\begin{equation}
L_{\rm FIR} = \,  4\,\pi\, R^2 \, \sigma \, T_d^{\,4} \ \ \ \ \ ,
\end{equation}
where $\sigma$ is the Stephan-Boltzmann constant. Similarly, the CO(1--0)  
luminosity 
\begin{equation}
\Lco = \,  4\,\pi^2\, R^2 \, 
{  {2\,k}  \over\lambda^2  } \int T_b\, d\nu \ \ \ ,
\end{equation}
where $T_b$ is the intrinsic brightness temperature, $\nu$ is frequency,
and $L_{\rm CO}$ is in the same units as $L_{\rm FIR}$ (e.g., Watts or \Lsun ).
We assume that regardless of how the gas is clumped in volume, the region has 
an area filling factor of unity.   If $T_b = T_d$, 

\begin{equation}
{ L_{\rm FIR} \over \Lco }  \ = \ 
{{ {\sigma\,T_d^{\,3}\,c^3}}
\over {2\,\pi\,k\,\nu_{\rm CO}^{\,3}\,f_V\,\Delta V}} \ \ \ , 
\end{equation}
where $\nu_{\rm CO}$ is the CO(1--0) rest frequency and 
$\Delta V$ is the linewidth.  For typical values,
\begin{equation}
{ L_{\rm FIR} \over \Lco  }   \ = \
4.8\times 10^6\, \left( {T_d \over 50K} \right)^{\,3} \ 
\left( { {300 }  \over {f_V\,\Delta V}} \right) \ \ \ .
\end{equation}

\begin{equation}
{ L_{\rm FIR} \over \Lco  }   \ = \
4.8\times 10^6\, \left({ { L_{\rm FIR}}  \over {10^{12} \ L_\odot} 
}\right)^{3/4}
\ \left({ {300 \ {\rm pc}} \over {R}} \right)^{3/2}
\ \left({ {300} \over {f_V\,\Delta V}} \right)  \ \ \ .
\end{equation}


With the CO luminosity in K\,\kms\,pc$^2$,
\begin{equation}
\LprimeCO \equiv  \pi\, R^2\,  T_b\, f_V\,\Delta V \ \ \ ,
\end{equation}
\noindent
the ratio becomes 
\begin{equation}
{ L_{\rm FIR}\over \LprimeCO }  \ = \
{  {4\,\sigma\,T_d^{\,4}} \over {f_V\,\Delta V \, T_b}  } \ \ \ .
\end{equation}
If $T_b = T_d$, then for typical values, 
\begin{equation}
{L_{\rm FIR} \over \LprimeCO}\ = \ 224\,\left( {T_d \over 50K}\right)^{\,3} \ 
\left( {300} \over  {f_V\ \Delta V}\right)  \ \ \ ,
\end{equation}
where the \Lfir /$L^\prime_{\rm CO}$ ratio now has units of \Lsun 
(\Kkmspc)$^{-1}$.

{\bf Figure~5} shows this predicted ratio vs.\ dust temperature,
for a typical $f_V\,\Delta V = 300$\,\kms , together with  
the observed ratios for ultraluminous galaxies in our sample.  
The median $L_{\rm FIR}/\LprimeCO $ = 160~\Lsun/\Kkmspc.  This is within
a factor of three of the black body limit for temperatures 
40 -- 80\,K.  Galaxies with $L_{\rm FIR} > 10^{12}$\,\Lsun\ 
are closer to the black body limit than the somewhat
less luminous galaxies in our sample.

Normal spirals have $L_{\rm FIR}/\LprimeCO$ ratios 10 to 30 times 
lower than the black body limit in {\bf Fig.~5}. In normal spirals, the 
gas and dust extend over several kpc, so the column density 
is lower and the dust is transparent to its own radiation in the far~IR.  The 
black body limit applies if the same amount of matter is confined to a 
smaller volume, making the 
dust opaque and allowing {\it all} the dust to  equilibrate with the radiation
field.   Ultraluminous galaxies indeed have $L_{\rm FIR}/\LprimeCO$ ratios 
close to the black body limit, showing their far~IR dust radiation is opaque  
at $\lambda \leq 100\,\mu$m. The absence of strong near-mid IR
(25$\mu$m) peaks in 90\% of our sample galaxies shows
opaque dust blocks our view into warmer components where the spectrum peaks in 
the near IR, so we see only an enormous, 
$\sim$300\,pc photosphere at $\sim 60$\,K.  

In reality, ultraluminous galaxies may differ from the black body model
in the following respects:

{\it a) CO brightness temperature $\not=$ dust temperature.}  At the mean 
H$_2$ density in the central regions of ultraluminous galaxies (typically 
$\sim 10^3-10^4$\,cm$^{-3}$), the gas and dust are probably not coupled, so the 
gas kinetic temperature may be only about half the dust temperature.  At 
that density, however, the {\it brightness} temperature of the CO(1--0) 
line will be close to the {\it gas} kinetic temperature. Although we took 
$T_b = T_d$ in our example, the argument is the same if the CO brightness 
temperature is half the dust temperature; the predicted 
$L_{\rm FIR}/\LprimeCO$ is twice as large, that is, 
even higher than the observed values ({\bf Fig.~5}).

{\it b) CO size larger than far~IR size.}
The gas probably occupies a larger volume than the opaque dust, 
which may be concentrated toward H~II  regions.  If so, the radius in eq.(5) 
will be larger than that in eq.~(4), lowering the predicted FIR/CO 
luminosity ratio toward the observed data ({\bf Fig.~5}).  Hence, a larger CO size 
compensates for a lower CO temperature, so the simple black 
body model may give the right ratio after all.

{\it c) Area filling factors.}
We assumed area filling factors of unity, but real sources have windows 
through the opaque dust that allow us to see near IR radiation and, in 
some sources, optical lines (redshifts of these objects are measured from 
optical lines, some of which, however, may come from gas outside of the opaque
nuclear regions).  Provided the area filling factors are about the same in 
the far~IR and in CO, the black body argument will be the same, and the 
predicted far~IR to CO ratio will be as in the above equations.  
\section{CO SIZE AND MOLECULAR MASS}
\subsection{Radii Predicted by the Black Body Model}
The black body equation (4) sets a lower limit to the far~IR source size.  
In a more complex model with several optically thick, massive sources, 
the true radius would be larger, but the radiating surface area 
would remain the same.  For some ultraluminous galaxies, the minimum radius is
about the same size as the observed nonthermal  radio source 
(Condon et al.\ 1991).  This far~IR
black body radius  ($R_{\rm bb}$ in {\bf Table~3}) may also be compared to the 
minimum CO radius derived from eq.\ (9) with the assumptions that the dust and 
CO brightness temperatures are the same and the velocity filling factor
$f_V$ is  unity.  

\begin{equation}
R_{\rm CO}({\rm min}) 
	= L^\prime_{\rm CO}/(\pi T_{\rm bb} \Delta V)^{0.5}\ \ \ \ .
\end{equation}
This is the minimum radius a CO source would have if its 
brightness temperature does not exceed the black body dust temperature.
For the sources in our sample, the minimum CO radius ({\bf Table~3}) is typically 
twice as large as the far~IR black body radius, although sometimes the two agree.
If these conditions are not fulfilled 
(e.g., a core-halo source), then the CO core radius scales as:
\begin{equation}
R_{\rm CO}({\rm core})= R_{\rm CO}({\rm min}) 			
{\left( 1 			\over f_V 	\right)^{0.5} }	   
{\left( L^\prime_{\rm core}	\over L^\prime	\right)^{0.5} }  
{\left( T_{\rm bb}		\over T_b	\right)^{0.5} }
\end{equation}
The minimum observed linewidths ({\bf Table~1}) are $\sim 150$\,\kms , presumably
from the face-on galaxies, while the larger observed linewidths suggest
a true rotation velocity $V\sim 300$\,\kms .  The velocity filling factor 
$f_V$ would thus be 150/300 $=$ 0.5 for edge-on galaxies, and 1.0 for face-on 
galaxies.  Interferometer observations of a few ultraluminous galaxies 
(e.g., Scoville et al.\  1991) suggest ($L^\prime_{\rm core}/ L^\prime ) 
= 0.5$, and $T_{\rm bb}/T_b = 2$. Hence on average, the true CO core
radii may be $\sqrt 2$ times larger than the minimum CO core radii listed in
{\bf Table~3}.
\subsection{Dynamical Mass as an Upper Limit on Gas Mass}
We calculated dynamical masses for the central regions of the ultraluminous 
galaxies by using the black body dust temperature derived from the 
60/100\,$\mu$m flux ratio to obtain a radius for the CO emitting region with
eq.(12).  We assume the characteristic velocity $V$ in this region is 
determined by gravity, so the enclosed dynamical mass is 
$R_{\rm CO}\, V^2\,/\,G$, or
\begin{equation}
M_{\rm dyn} = \left({L_{\rm CO}\over {\pi\,T_b\,f_V\,\Delta V}}\right)^{0.5} 
{ V^2 \over G} \ \ \ \ \ .
\end{equation}
For a velocity filling factor $f_V = 1$, 
the CO radius is a minimum radius, so a dynamical mass estimated this way 
is also a minimum value.  As above, $T_b$ is the CO intrinsic brightness 
temperature (assumed here to be the black body dust temperature), $\Delta V$
is the observed CO linewidth ({\bf Table~1}), and $V$ is the true velocity in the region.

In our sample, galaxy inclinations are unknown, so we estimated the true 
velocity $V$ statistically. The histogram of CO linewidths for the sample 
({\bf Fig.~6}) has a median of 300\,\kms.   We therefore assumed all
the galaxies in our sample had a minimum internal velocity
of 300\,\kms , and that those with smaller observed linewidths were 
inclined to our line of sight.
The face-on galaxies are those with the minimum observed linewidths 
of $\sim 150$\,\kms.  To calculate minimum dynamical masses listed in 
{\bf Table~3}, we thus took the minimum CO core radii, with $f_V = 1$ in eq.(9), 
and we took $V= {\rm max} (300$\,\kms\ , $\Delta V$).  If the motions are primarily radial, then
${\rm V} = 3^{0.5}\, \Delta V$.

We also list in {\bf Table~3} the equivalent total H$_2$ density, $n_{\rm tot}$,  
a maximum to the true volume-averaged \htwo\ density,   derived from
the dynamical masses and the minimum CO radii.   
To calculate the minimum CO radii, we assumed
thermalization, with the CO(1--0) brightness temperature 
equal to the gas kinetic temperature.  At a kinetic temperature of 60\,K,
this would occur for H$_2$ densities $>1500$\,cm$^{-3}$.  
Even if the CO emitting gas were to evenly fill the volume inside $R_{\rm CO}$ 
as an intercloud medium, the CO rotational
levels would be thermalized in about half of the sources.

Since the gas mass, $M$(H$_2$), cannot exceed the dynamical mass, the
average ratio $M_{\rm dyn}/L^\prime_{\rm CO} \, = 1.4$ ({\bf Table~3}) 
suggests that $\alpha$, 
the H$_2$-mass-to-CO-luminosity ratio, may be at least 
three times lower than in Galactic molecular clouds.
This is consistent with the black body model.   
For gravitationally bound clouds,  this ratio is 
$\alpha = 2.1\cdot n^{0.5}/T_b$ (e.g., Radford, et al.\ 1991a); CO 
thermalization  requires that 
$n$(H$_2) = 2000$\,cm$^{-3}$ to obtain $T_b = 60$\,K, yielding 
$\alpha \, \approx \,1.6$\,\Msun /\Kkmspc . 

The dynamical mass listed in {\bf Table~3} is the minimum dynamical 
mass of a core source that has {\it all} the observed CO luminosity, 
$L^\prime_{\rm CO}$, a CO brightness temperature $T_b$(CO) equal to the 
far~IR black body temperature $T_{\rm bb}$, and a velocity filling factor 
$f_V = 1$.  If these assumptions are not fulfilled (e.g., a core-halo 
source), then  dynamical mass scales linearly with $R_{\rm CO}$(core) as in 
eq.(13). Hence to get the true dynamical mass in the CO core 
region, the values of $M_{\rm dyn}({\rm min})$ in {\bf Table~3} should be 
multiplied by 
\begin{equation}
{M_{\rm dyn}({\rm core}) \over M_{\rm dyn}({\rm min})} 		=
{\left( 1 			\over f_V 	\right)^{0.5} }	
{\left( L^\prime_{\rm core}	\over L^\prime	\right)^{0.5} }  
{\left( T_{\rm bb}		\over T_b({\rm CO})	\right)^{0.5} } .
\end{equation}
Similarly, the ratio
${M_{\rm dyn}({\rm min})/ L^\prime_{\rm CO}}$ listed in {\bf Table~3}
can be scaled by the same factor to get
the true ratio of core dynamical mass to total CO luminosity:
\begin{equation}
{M_{\rm dyn}({\rm core})\over L^\prime_{\rm CO}} = 
{M_{\rm dyn}({\rm min}) \over L^\prime_{\rm CO}} 		
{\left( 1 			\over f_V 	\right)^{0.5} }	
{\left( L^\prime_{\rm core}	\over L^\prime	\right)^{0.5} }  
{\left( T_{\rm bb}		\over T_b({\rm CO})	\right)^{0.5} } .
\end{equation}
For the values suggested above, namely, $f_V$ = 0.5, 
($L^\prime_{\rm core}/ L^\prime ) = 0.5$, and 
$T_{\rm bb}/T_b({\rm CO}) = 2$, the true dynamical mass and the ratio 
$M_{\rm dyn}/L^\prime$ would both be $\sqrt 2$ times higher than the values 
in {\bf Table~3}.

A check on the dynamical masses derived here can be provided by the
fits to the 2.3\,$\mu$m bandhead absorption originating in stellar 
atmospheres to estimate the velocity dispersion of the nuclear bulge
stars.  In a 1.5$''$ square aperture centered on NGC\,6240S, 
Lester \& Gaffney (1994) find $\sigma = 350$\,\kms , and estimate 
$M(<260$\, pc)$ = 2\ R \ \sigma^2/G = 1.9 \times 10^{10}$\,\Msun .  This 
agrees reasonably well with  our measurement  ({\bf Table~3}),  of $\Delta V = 370$\,\kms , and
$M(<340$\, pc)$=R\ \Delta V^2/G = 1.1 \times 10^{10}$\,\Msun .
A similar estimate has been made for Arp\,220 by Shier, Rieke, \& Rieke
(1994), yielding $M(<350$\,pc)$=1.8 \times 10^9$\,\Msun , which is 
ten times lower than our estimate of the dynamical mass ({\bf Table~3}).  This
is mainly due to the low velocity dispersion of only 125\,\kms\ deduced
by Shier et al., whereas we took the observed Arp\,220 CO linewidth of 
480\,\kms .   The large CO linewidth is also present in our millimeter
spectra of HCN and CS. We suspect that the 2\,$\mu$m CO bandhead data from 
Arp\,220 may be affected by the heavy extinction to the western near IR peak, 
and may not refer to the same volume as the millimeter line data. 
\subsection{Lower Limit to \htwo\ Mass: \ Optically Thin CO}
Since CO is always optically thick in Milky Way GMC's and spiral galaxies, 
we can get a useful lower limit to the \htwo\ mass by
assuming the CO(1--0) line has an optical depth $\tau \leq \,1.0$.  
The minimum H$_2$ gas mass is then 
\begin{equation}
M_{\rm thin} = 6.9\times 10^{-3}\ M_\odot \ \Lcop \  \ T_b
\end{equation}  
for \Lcop in \Kkmspc and $T_b$ in K.
We assume the rotational levels are thermalized and the CO/\htwo\  abundance ratio 
is  $1.0 \ \times \ 10^{-4}$.    The black body model for
$T_b$ and the observed line luminosities 
yield an optically thin  molecular mass given in   {\bf Table~4}. 
They are typically 2 -- $5 \times10^9$~\Msun , or about one-third the 
dynamical mass.   Even if the emission is optically thin  a large fraction of 
the dynamical mass is H$_2$  with 
$M_{\rm thin} = \, 0.34 \pm \ 0.12 \, M_{\rm dyn}$.   
 Evidence from CO excitation points to optical depths $\tau \ > \
\,1.0$\ (Radford, et al.\ 1991a) indicating that a mass based on
 the optically thin approximation is good lower limit. 
In this model, if the CO optical depth is only $\sim 3$, then
molecular gas alone would account for most of the dynamical mass.
While the assumption of thermalized rotational levels may 
increase the thin mass estimate by a factor of 2 due to an overestimate of 
the partition function, this will be more than compensated for by the 
realistic optical depth.  The {\it minimum}
(optically thin)  molecular mass would be lower only if the metallicity were
significantly greater than Solar and the CO/\htwo\  abundance ratio  much greater than
in Milky Way GMC's. 

\subsection {Dust Mass from 100 \micron\ Emission}

An alternative method of estimating the mass of molecular gas which is independent of
spectral line observations is to use the far infrared emission and assume
the dust is optically thin. This is, of course, inconsistent with the black 
body model of optically thick dust but it provides an alternative measure 
of mass. To calculate the dust mass we assume the dust is thin, its emissivity
varies as $\lambda  ^{-1.5}$, and its far IR
emission coefficient is that given by Hildebrand (1983). This should be 
regarded as a rough estimate since IR emissivity of dust in molecular clouds 
is not well calibrated.  The dust temperature  determined from the 
60/100 \micron \ ratio is typically 40 -- 50 K, and the
mass is estimated from the 100 \micron  \ flux.  Since this measures only 
warm dust radiating at 100
\microns, \ it is strictly a lower limit to the total dust mass. 
Near the center of a galaxy, however, most of the dust will indeed be warm.  
If we further assume the gas-to-dust 
mass ratio is 100, the total molecular mass can be calculated  
(column 4 of {\bf Table~4}).  The average
gas mass estimated from the  warm dust emission is very close to the dynamical
mass, with $100 \,M_{\rm dust} \, = \, 1.1 \pm 0.5 \, M_{\rm dyn}$

\subsection{Modified H$_2$ mass-to-CO luminosity Relation}
For a galaxy containing virialized molecular clouds, the 
H$_2$ mass-to-CO luminosity relation can be expressed as 
\begin{equation}
M^\prime({\rm H}_2) = \alpha L^\prime_{\rm CO} .
\end{equation}
Use of the Milky Way H$_2$ mass-to-CO luminosity
 ratio, $\alpha = 4.6$\,\Msun /\Kkmspc,  
clearly overestimates the \htwo\ mass in ultraluminous galaxies ({\bf Table~2}). 
We previously (DSR) related the CO luminosity to the dynamical mass as well 
as the \htwo\,\ mass.  
This model may explain why the true gas masses are lower than the 
masses derived with the Milky Way ratio.  The key point is that unlike
Galactic clouds or gas distributed in the disks of galaxies, the CO in the 
centers of ultraluminous galaxies may {\it not} come from virialized
clouds, but from a filled intercloud medium,  so 
{\it the linewidth is determined by the total dynamical
mass in the region (gas and stars)}, that is, $\Delta V^2 = G\, \Mdyn /R $. 
The CO line emission may trace a medium bound by the total 
potential of the galactic center, containing a mass \Mdyn\ consisting of 
stars, dense clumps, and an
interclump medium containing the CO emitting gas with mass $M$(H$_2$).
 
Defining $f \equiv M({\rm gas})/M_{\rm dyn} $,
we showed the usual CO to \htwo\  mass relation becomes
\begin {equation}
\displaystyle
M_{\rm dyn}/\LprimeCO  \ = \ f^{-1/2} \,\alpha\, 
		\ = \ f^{-1/2} \,2.6\,(\nbarH2)^{1/2}\,T_{\rm b}^{-1} 
\ \ ,
\end{equation}
\begin {equation}
\displaystyle
 M({\rm H}_2)/\LprimeCO  \ = \ f^{1/2} \,\alpha\,
		\ = \ f^{1/2} \,2.6\,(\nbarH2)^{1/2}\,T_{\rm b}^{-1} 
\ \ ,
\end{equation}
and 
\begin{equation}
\displaystyle
\Mdyn  M({\rm H}_2)  = (\alpha\,\LprimeCO)^2 
\ \ ,
\end{equation}
where $\nbarH2$ is the H$_2$  density averaged over the whole volume.
We argued in our earlier paper that the quantity $\alpha \LprimeCO$  
measures the geometric mean of total mass and gas mass.  It
underestimates total mass and overestimates gas mass.
Hence if the CO emission in ultraluminous galaxies comes from
regions not confined by self gravity, but instead from an intercloud
medium bound by the potential of the galaxy, or from molecular gas
in pressure, rather than gravitational, equilibrium, then the usual relation 
$M({\rm H}_2)/$\LprimeCO $=\alpha$ must be changed to eq.(20).

\subsection{Summary of Revised Mass Estimates}
For the compact central regions, the dynamical mass is the best estimator 
of the molecular mass. The dynamical masses listed in {\bf Table~4} are those for the black-body
model; the actual dynamical mass is given by eq.(15). We have argued here that the
black-body model is a good approximation for ultraluminous galaxies since it comes close
to predicting the ratio of far IR to CO luminosity.   The  CO luminosity  to molecular mass
conversion factor  for ultraluminous galaxies  must be $\leq$  
$M_{\rm dyn}$/\LprimeCO\ $\sim1.4$. is given in {\bf Table~3}). 
 Our estimates of the  gas mass from optically thin CO(1-0)  (Column 3 of 
{\bf  Table~4})  and from
optically thin dust (Column 4 of {\bf Table~4}) indicate that a large fraction of the dynamical mass
is molecular gas (i.e., $f$ approaches 1 in equations 19 and 20).  If the dynamical mass is dominated
by the gas mass  and the core size is close to that of the black-body model then 
\begin{equation}
 M({\rm H}_2)  = 1.0 \times 10^{10} M_\odot,
\end{equation}
with a dispersion of only 0.3 $\times 10^{10}$\,\Msun .  
This is about three times lower than previous estimates of the gas mass 
which have utilized the Milky Way $M({\rm H}_2)/$\LprimeCO\ ratio. Even if the velocity filling
factor is  0.5 and the excitation temperature half that of the dust the dynamical mass is only
increased by a factor of 2.  More accurate determinations of the
dynamical mass will come from interferometer measurements  of the CO radii  for the
nearer galaxies in the survey.

   The standard 
M$^\prime({\rm H}_2)$/\LprimeCO\ = 4.6\,\Msun /\Kkmspc  (Column 2 of
{\bf Table~2}),  
appropriate for the  molecular gas in a more typical galactic
environment outside the central  core, is  an overestimate for 
ultraluminous galaxies.

\section{PAIR SEPARATION AND \Lfir/\Lco  OF ULTRALUMINOUS GALAXIES}
The prevalence of close interactions among IR bright galaxies was noted 
soon after the first IRAS galaxy catalog became available (Lonsdale,
Persson, \& Matthews 1984; Soifer et al.\  1984).
In  {\bf Fig. 7} we present R-band CCD images of 12 sample galaxies between RA 10h to
23h (Sanders \& Kim, private communication);  they 
show a wide variation in morphology  and pair separation.
Some appear distinctly single, with only slightly distorted disks, 
e.g., 1609--0139, while others are prominent, closely interacting galaxies 
with two disks and two nuclei, e.g., 10190+1322 or 15030+4835.  One of the most luminous and
most distant objects in our  sample, 14070+0525, appears to be an isolated galaxy,
although very faint  tails would be invisible on this image.
{\bf Table 5} summarizes the data on pair separation for the  sample, from the literature
and from {\bf Fig. 7 }.  The measured separations range from 0.3 
to 14 kpc with an average of 6 kpc.    Near IR images of  some close 
ultraluminous galaxies such as Arp\,220 and Mrk\,231 show that   galaxies apparently
single in optical images may have double nuclei with small separations,  indicating a
nearly completed merger (Graham et al.\  1990; Majewski et al.\  1993;  Armus et al.\ 
1994).   Conversely some galaxies which appear double in optical images clearly have
only one nucleus in K band images (Murphy et al.\ 1996).  Given the large amount of dust
in these objects, however,  even the near IR images may be heavily affected by
extinction, and suspected double nuclei may actually be bright objects outside of 
 the heavily obscured CO-emitting zone.

   On the basis of R band images
of ten nearby ultraluminous  galaxies, Sanders et al.\  (1988a) argued  that all
ultraluminous galaxies are  strongly interacting, as shown by tidal tails,  rings or 
double nuclei with separations
$< 5$\,kpc, and  that the fraction of close doubles  increases with IR luminosity.  From
optical morphology and spectra,  they suggested ultraluminous galaxies represent the
initial dust  enshrouded stage in the formation of quasars. In their model, most of the 
luminosity comes from an AGN rather than star formation.  Because very few of the
ultraluminous galaxies  found by IRAS have true quasar or Seyfert~1 spectra, Sanders et
al.\  argued  the AGN was completely hidden by dust and the  transition  phase to a
dust-free AGN is represented by the \char'134 warm"  ultraluminous galaxies with
$S_{25}/S_{60} > 0.2$ (Sanders et al.\  1988b). Only five of the objects in our sample  have
$S_{25}/S_{60} > 0.2$ and four of these  overlap with Sanders et al.'s\  sample.   However,
the object with the highest
\Lfir/$L^\prime_{\rm CO}$,  08572+3915, is a warm galaxy but is definitely not a
completed merger  since its separation is 5.5\,kpc (Sanders 1992).

There are no significant correlations of pair separation (Table 5) with 
CO luminosity or 
more importantly the ratio of far IR to CO luminosity,  \Lfir/$L^\prime_{\rm CO}$ .
  The apparent stage of
the interaction appears to have  little effect on the efficiency of star formation or the
fueling of the  putative black hole.   Extremely luminous infrared galaxies include completed
mergers and interacting pairs with separations of approximately 10kpc.  Two of the more
distant  and more luminous  objects in our sample, 16334+4630 and 15030+4835, with
very cool   far~IR colors, have large pair separations, 11 and 
12 kpc ({\bf Fig.~7}) respectively. Both have
\Lfir \ $\approx 2 \times 10^{12}$\Lsun , similar to Mrk\,231 or the quasar 
Mrk\,1014.   If an
evolutionary sequence is present in ultraluminous galaxies it is not apparent from  the CO
or IR data combined with the optical morphology. 

\newpage
\section{DISCUSSION}

\subsection{ Dense Molecular Gas}
In addition to a high IR/CO luminosity  ratio,
the  molecular gas in the centers of ultraluminous galaxies exhibits  other
  differences from  GMC's  in the Milky Way and normal spiral galaxies.  
Most important
is the  extremely high HCN luminosity. 
  HCN emission traces H$_2$ at a much higher 
density, $\sim 10^5$~cm$^{-3}$, than CO \hbox{($\sim500$~cm$^{-3}$)}.  
As we have shown previously, the HCN luminosities of the ultraluminous galaxies
Mrk~231, Arp~220, and NGC~6240 are greater than the CO luminosity of the Milky 
Way (Solomon, Downes, \& Radford 1992a).    The ratio of HCN to 
CO luminosity is 1/4 to 1/7 for ultraluminous galaxies, but only 1/30 
or less in the disks of normal spiral galaxies.  This implies that a 
large fraction of the molecular gas in ultraluminous galaxies, perhaps 
50\%, is in very dense regions similar to star forming cloud cores in 
Orion or W51, 
rather than in the envelopes of giant molecular clouds, as is the case in the 
disks of normal spiral galaxies.

 If galactic HCN luminosity is taken as a
dense gas indicator and far infrared luminosity as a high mass star
 formation indicator,
then the star formation rate per mass of  {\it dense} gas is the same for normal spirals
and IR luminous interacting galaxies.  There is    sufficient dense gas in ultraluminous galaxies
to account for their luminosity by star formation.    
 In our model, the CO emission from ultraluminous galaxies
originates from an intercloud medium essentially filling the volume;
the emission from HCN and CS is from the dense ``clouds'' embedded inside the
molecular region. The entire ISM is a scaled up version of a normal galactic 
disk. This environment is an ideal stellar nursery for
prodigious star formation leading to ultraluminous galaxies.

\subsection{Summary of CO Observations and \htwo\,$\,$Mass }

Of the 37 galaxies in the sample, 32 have a far infrared luminosity greater than 6.5 x 
$10^{11}$ \Lsun and total IR luminosities greater than  8 x 
$10^{11}$ \Lsun. We refer to these as
ultraluminous IR galaxies.  They share several properties in common:

\begin{enumerate}  

\item  All but one of the ultraluminous galaxies
have a high CO(1--0) luminosity, with values of 
log (\Lprimeco /\Kkmspc ) $= 9.92\, \pm 0.12 $ . The extremely small
dispersion of only 30 \% is half that of the far infrared luminosity. 
The
parameter with the narrowest range (best defined) is not IR luminosity but CO
luminosity, even though the selection was based on IR luminosity.

\item 
We have demonstrated that for ultraluminous galaxies in which the central
few hundred parsecs is dominated by a molecular core, the CO luminosity and the
ratio of FIR to CO luminosity can be explained as a consequence of thermal 
emission from optically thick dust and CO
in  a central sphere, torus (ring) or disk.
 The ambient densities are a factor of 100 higher than in a galactic disk, making even the
intercloud medium molecular throughout. 
  The median value of
 $L_{\rm FIR}/\LprimeCO $ = 160~\Lsun/\Kkmspc , within
a factor of 2   of the black body limit for the observed far IR temperatures.

\item From the observed linewidths, we estimate
the dynamical mass within the minimum CO radius ($\sim $ 400 pc ).  
This provides an upper limit on the mass of molecular gas. For a velocity filling
factor of 1.0, the dynamical mass is 1.0 $\pm 0.3 \times 10^{10}$\Msun .  A more likely velocity
filling factor of 0.5  raises the upper limit to the mass by $2^{0.5}\,$. 

\item We assume the CO is optically thin in the (1--0) line to estimate the 
gas mass from the observed CO luminosity.  This provides   
a lower limit on the mass of molecular gas of typically 
2 -- $5 \times10^9$~\Msun , with 
$M_{\rm thin} = \, 0.34 \pm \ 0.12 \, M_{\rm dyn}$
or about one-third the dynamical mass. 

\item We assume the dust is optically thin at 100\,\microns\ to 
 estimate the minimum mass of interstellar material.  This provides another,
 independent  measure of the mass of molecular gas.  The average gas
 mass estimated from the dust continuum emission assuming a dust to gas ratio of 100 
is very
close to the dynamical
 mass,  
 $100 \,M_{\rm dust} \, = \, 1.1 \pm 0.5 \, M_{\rm dyn}$.

\item
The standard  Milky Way ratio, 
$M^\prime({\rm H}_2)$/ \LprimeCO\ = 4.6  \Msun (\Kkmspc )$^{-1}$, 
yields  a mass   $M^\prime({\rm H}_2) = \, 3.7\, M_{\rm dyn}$,  which is
clearly  an overestimate. We show that in the  extreme environment near the center of an
ultraluminous galaxy, where the CO emission originates from an intercloud medium which
essentially fills a volume rather than from clouds bound by self gravity,  the CO luminosity traces
the geometric mean of the molecular mass and the total dynamical  mass.  The true total molecular
mass must be between the optically thin CO estimate and the dynamical mass.  The agreement
between the total gas mass estimated from the dust emission and the dynamical mass suggests a
number closer to the dynamical mass.   
 Dynamical
mass estimates with sizes determined from interferometer measurements  are required to
obtain better estimates of the molecular mass in ultraluminous galaxies.

\end{enumerate}

\newpage
\subsection{The Effect of Close Interactions on the Molecular ISM}

 Molecular clouds are in virial equilibrium, not
 pressure equilibrium, (eg. Solomon et al.,
1987)  maintained by a balance between self gravity and turbulence driven primarily by star
formation.    The external pressure from the ambient ISM or
intercloud medium is significantly less than the internal effective pressure
characterized by supersonic velocities. The star
formation is self regulating and GMC's in galactic disks have modest star formation
rates per solar mass ( Mooney and Solomon, 1988).

The uniformly high CO luminosities found here for all ultraluminous
 galaxies, combined with the evidence of very large quantities of even
denser molecular gas, show that the  molecular gas required for an extreme  starburst is
always  present when a galaxy has ultrahigh infrared luminosity. 
  Does this imply that a
merger or close interaction is the ideal site for the transformation of diffuse gas into
molecular clouds or the stimulation of high mass star formation  by cloud collisions
 (eg. Scoville, Sanders \& Clemens, 1986)?  Both seem highly unlikely given the
typical interaction velocity of 200 or 300 km/s, more than an order of magnitude greater
than the escape velocity of material in even the most massive GMC's, and more than
sufficient to completely  ionize and destroy any trace of molecules in a direct collision
between clouds.  The molecular gas in luminous mergers must be from {\bf preexisting} 
molecular clouds. Unlike diffuse gas, the mean free path for molecular clouds is
sufficiently long (
$\sim$ 2 kpc) that a merger, unless it is completely edge on, will result in few direct
collisions.  Molecular clouds can therefore survive a close
 interaction or merger, although
they will be changed by the new environment.

Two mechanisms have been proposed for triggering an intense burst of star formation in
preexisting molecular clouds during a close interaction or merger. During a direct collision the HI
clouds will collide, forming a hot ionized high pressure remnant gas
\hbox{( Jog and Solomon, 1992)}. The
overpressure due to this hot gas causes a radiative shock compression of the outer layers of GMC's
in the overlapping wedge region. The outer layers become gravitationally unstable, leading to a
burst of star formation in the initially stable GMC's.   This scenario probably applies to luminous 
($\sim 10^{11}$\lsun) IR galaxies with prominent extranuclear starbursts such as Arp 299, but is
not strong enough to produce ultraluminous galaxies. A more efficient mechanism for producing a
central starburst in interacting galaxy pairs from preexisting	 clouds has been suggested by Jog
and Das (1992).  In this model, as a disk GMC tumbles into the central regions of a galaxy following
an encounter, it undergoes  radiative shock compression by the  high pressure of the central
molecular intercloud medium. When the growth time for the gravitational instabilities in the
shocked outer shell of a cloud becomes smaller than the shock crossing time, the shell becomes
unstable, resulting in a burst of star formation.  The infall may be helped by the establishment of a
nonaxisymmetric bar potential (eg. Norman , 1991).  The luminosity depends on the compressed
mass fraction, cloud infall rate and efficiency of star formation; evolved mergers  generate a
luminosity comparable to that of ultraluminous IR galaxies.

 Our analysis of CO emission from ultraluminous
galaxies reduces the \htwo\,mass from previous estimates of 2 --5 $\times 10^{10}\msun$ to
0.4--1.0 $\times 10^{10}\msun$, which is in the range found for molecular gas rich spiral galaxies.
Thus a collision involving a molecular gas rich spiral could lead to an ultraluminous galaxy
powered by  central starbursts triggered by  the compression of  preexisting GMC's.

\acknowledgments
We thank Mr. Yu Gao, Stony Brook, for help with Table~5, and 
Fred Seward, Elizabeth Bohlen, 
and the Center for Astrophysics for the use of their measuring 
engine.   We especially thank D. B. Sanders and  D. C. Kim  for letting us use
their R band CCD images. 

\clearpage

\begin{planotable}{lcclcll}

\tablecaption{DATA FROM THE IRAM 30\,m
TELESCOPE ON CO(1--0) IN ULTRALUMINOUS GALAXIES}
\label{CO(1---0) Data on Ultraluminous Galaxies}
\tablehead{
\colhead{Source} &\colhead{Position} &\colhead{Observed} &\colhead{Redshift} 
		&\colhead{Linewidth} &\colhead{Intensity} &\colhead{CO/FIR}
\nl
\colhead{name}	&\colhead{R.A.\ 1950}	&\colhead{Dec. 1950} &\colhead{$cz$}
		&\colhead{$\Delta V$}	&\colhead{$I_{\rm CO}$}
                &\colhead{  $I_{\rm CO}/S_{100\mu{\rm m}}$  }  
\nl
\colhead{}	&\colhead{(\ \,h\ \ m\ \ s)}
		&\colhead{(\ \ \ $^\circ$\ \ \ $'$\ \ \ $''$)}
		&\colhead{(\,km\, s$^{-1}$)}
		&\colhead{(km\, s$^{-1}$)}		
		&\colhead{(K\, km\ts s$^{-1}$)}\phantom{X}
		&\colhead{(Kkm\ts s$^{-1}$\ts Jy$^{-1}$)}
}
\startdata
00057$+$4021 	&00 05 45.1	&$+$40 21 14 		&\phantom{6}13390
		&350		&\phantom{XX10}9.9	&2.30
\nl		
00188$-$0856 	&00 18 53.7	&$-$08 56 07		&\phantom{6}38530
		&390		&\phantom{XX10}2.2	&0.65
\nl		
00262$+$4251	&00 26 12.9	&$+$42 51 40		&\phantom{6}29153
		&230		&\phantom{XX10}3.5	&1.43
\nl
I Zw 1  	&00 50 57.8	&$+$12 25 19		&\phantom{6}18330
		&410		&\phantom{XX10}7.0$^{b)}$	&2.66
\nl		
Mrk\,1014 	&01 57 16.6	&+00 09 07		&\phantom{6}48947
		&200		&\phantom{XX10}1.8	&0.83
\nl
\nl
02483$+$4302 	&02 48 20.4	&$+$43 02 56		&\phantom{6}15419
		&250		&\phantom{XX10}5.7	&0.82
\nl		
03158$+$4227 	&03 15 52.3	&$+$42 27 36		&\phantom{6}40296
		&180		&\phantom{XX10}2.1 	&0.49
\nl		
03521$+$0028 	&03 52 07.8	&$+$00 28 20		&\phantom{6}45530
		&150		&\phantom{XX10}2.2	&0.57
\nl		
04232$+$1436 	&04 23 15.2	&$+$14 36 53		&\phantom{6}23855
		&400		&\phantom{XX10}7.5	&1.76
\nl		
VII\,Zw\,31 	&05 08 17.5	&$+$79 36 40		&\phantom{6}16260
		&200		&\phantom{XX1}21.0	&2.18
\nl		
\nl		
07598$+$6508 	&07 59 53.0	&$+$65 08 21	&\phantom{6}44621$^{a)}$
		&337		&\phantom{XX10}2.8$^{a)}$	&1.62
\nl
08030$+$5243 	&08 03 01.5	&$+$52 43 45		&\phantom{6}25031
		&420		&\phantom{XX10}6.7	&1.53
\nl
08572$+$3915 	&08 57 13.0	&$+$39 15 39		&\phantom{6}17450
		&270		&\phantom{XX10}2.0	&0.44
\nl
09320$+$6134 	&09 32 04.7	&$+$61 34 37		&\phantom{6}11785
		&350		&\phantom{XX1}15.6	&0.77
\nl		
10035$+$4852 	&10 03 35.5	&$+$48 52 25		&\phantom{6}19427
		&250		&\phantom{XX10}8.8	&1.41
\nl
\nl		
10190$+$1322 	&10 19 01.4	&$+$13 22 04		&\phantom{6}22953
		&390		&\phantom{XX10}7.4	&1.33
\nl		
10495$+$4424 	&10 49 30.1	&$+$44 24 46		&\phantom{6}27674
		&330		&\phantom{XX10}5.1	&0.94
\nl		
10565$+$2448 	&10 56 35.4	&$+$24 48 43		&\phantom{6}12923
		&300		&\phantom{XX1}15.7	&1.04
\nl		
\hbox{11506$+$1331} 	&11 50 39.8	&$+$13 31 05		&\phantom{6}38158
		&290		&\phantom{XX10}2.5	&0.75
\nl		
Mrk\,231	&12 54 05.0	&$+$57 08 39		&\phantom{6}12650
		&230		&\phantom{XX1}22.0	&0.73
\nl		
\nl		
13106$-$0922 	&13 10 37.3	&$-$09 22 15		&\phantom{6}52290
		&200		&\phantom{XX10}1.7	&0.70
\nl		
Arp\,193 	&13 18 17.0	&$+$34 24 07		&\phantom{60}7000
		&410		&\phantom{XX1}36.0	&1.43
\nl		
Mrk\,273 	&13 42 51.6	&$+$56 08 14		&\phantom{6}11324
		&300		&\phantom{XX1}19.0	&0.89
\nl		
13442$+$2321 	&13 44 18.0	&$+$23 21 14	&\phantom{6}42620\phantom{-XX}
		&140		&\phantom{XX10}1.4	&0.62
\nl
14070$+$0525	&14 07 00.5	&$+$05 25 41		&\phantom{6}79621
		&270	&\phantom{XX10}0.8\phantom{XXXXX}&0.42
\tablebreak
\nl
15030$+$4835 	&15 03 01.3	&$+$48 35 24	&\phantom{6}64900\phantom{-XX}
		&270		&\phantom{XX10}1.5 	&1.03
\nl		
Arp\,220 	&15 32 46.9	&+23 40 08		&\phantom{60}5450
		&480		&\phantom{XX}109.0			&0.97
\nl		
16090$-$0139	&16 09 04.9	&$-$01 39 25		&\phantom{6}40044
		&300		&\phantom{XX10}3.7	&0.76
\nl
16334$+$4630	&16 33 24.3	&$+$46 30 58		&\phantom{6}57250
		&320		&\phantom{XX10}1.5	&0.69
\nl		
NGC\,6240 	&16 50 27.2	&+02 28 58		&\phantom{60}7298
		&370		&\phantom{XX1}69.0 	&2.48
\nl
\nl		
17208$-$0014 	&17 20 48.2	&$-$00 14 17		&\phantom{6}12837
		&360		&\phantom{XX1}20.0	&0.57
\nl
18368$+$3549 	&18 36 49.5	&$+$35 49 36		&\phantom{6}34850
		&330		&\phantom{XX10}3.4	&0.89
\nl		
19297$-$0406	&19 29 43.1	&$-$04 06 24		&\phantom{6}25700
		&300		&\phantom{XX10}6.8	&0.88
\nl		
19458$+$0944	&19 45 52.0	&$+$09 44 30		&\phantom{6}29980
		&350 		&\phantom{XX10}6.4 	&0.90
\nl		
20087$-$0308 	&20 08 46.4	&$-$03 08 52		&\phantom{6}31688
		&400		&\phantom{XX10}7.7	&1.18
\nl
\nl		
22542$+$0833 	&22 54 11.3	&$+$08 33 22		&\phantom{6}49750
		&150		&\phantom{XX10}1.2	&0.81
\nl		
23365$+$3604 	&23 36 32.3	&$+$36 04 34		&\phantom{6}19330
	&310	&\phantom{XX1}10.7\phantom{XXXXX}	&1.28
\nl
\nl
{\it median:}	& ---		& ---			&
		& 300		&\phantom{XX10} --- 	&0.9
\nl
{\it std. dev.:}& ---		& ---			&
		&\phantom{1}85	&\phantom{XX10} ---	&0.6
%
\tablenotetext{}{
$^{a)}$Sanders et al.\  (1988b), $I_{\rm CO}$ corrected by us;
$^{b)}$Barvainis et al.\  (1989).
}
\tablenotetext{}{
Errors:  $cz$: $\pm\ts$20 km\ts s$^{-1}$; 
$\Delta V$: $\pm\ts$30 km\ts s$^{-1}$; 
$I_{\rm CO}$: $\pm\ts$20\% for $I>$6, \  $\pm\ts$35\% for $2<I<$6.
}  
\tablenotetext{}{
Linewidths are Gaussian fit values (FWHP), or else half the full width to zero 
intensity} 
\tablenotetext{}{
for non-gaussian lines (See {\bf Figure~2} for spectra).
}  
\tablenotetext{}{
For point sources, $S/T_{\rm mb} = 4.5$\, Jy/K at the 30\,m telescope at 3\,mm.
}
\end{planotable}
\clearpage
\begin{planotable}{llllllllcc}
\tablecaption{CO and far~IR Luminosities of Ultraluminous Galaxies}
\label{CO and far~IR Luminosities of Ultraluminous Galaxies}
\tablehead{
\colhead{Source}&\colhead{\Lcop }	&\colhead{$M^\prime({\rm H}_2)$} 
		&\colhead{$S_{25}$}	&\colhead{$S_{60}$}	
		&\colhead{$S_{100}$}	
&\colhead{$\displaystyle \strut { S_{60} \over \displaystyle \strut S_{100} }$}
                &\colhead{$L_{\rm FIR}$}
&\colhead{\phantom{XX}$\displaystyle \strut {L_{\rm FIR}\over 
		\displaystyle \strut M^\prime({\rm H}_2)}$}
&\colhead{\phantom{XX}$\displaystyle \strut {L_{\rm FIR} \over 
 \displaystyle \strut L^\prime_{\rm CO} }$ }  
\nl
\colhead{name}	&\colhead{(10$^9$\,L$_l$)}	&\colhead{(10$^{10}$\,\Msun )}
		&\colhead{(Jy)}	&\colhead{(Jy)}	&\colhead{(Jy)}	&\colhead{}
		&\colhead{(10$^{12}$\,\Lsun )}	
&\colhead{\phantom{XX}$\displaystyle \strut {\rm L}_\odot 
	\over \displaystyle \strut {\rm M}_\odot $}
&\colhead{$\displaystyle \strut {\rm L}_\odot 
	\over \displaystyle \strut {\rm L}_l $}
}
\startdata
00057$+$4021 	&\phantom{XX1}3.8\phantom{XX}		&1.7 	
	&\phantom{$<$}0.36	&\phantom{10}4.47	&\phantom{11}4.30
	&1.04		&0.31	&\phantom{1}17		&\phantom{7}80
\nl		
00188$-$0856 	&\phantom{XX1}6.7		&3.1 	
	&\phantom{$<$}0.37	&\phantom{10}2.59	&\phantom{11}3.40
	&0.76		&1.51	&\phantom{1}48		&224
\nl		
00262$+$4251	&\phantom{XX1}6.2		&2.9 	
	&\phantom{$<$}0.33	&\phantom{10}2.98	&\phantom{11}2.44
	&1.22		&0.99	&\phantom{1}34		&160
\nl
I Zw 1  	&\phantom{XX1}5.0		&2.3 	
	&\phantom{$<$}1.21	&\phantom{10}2.24	&\phantom{11}2.63
	&0.85		&0.29	&\phantom{1}12		&\phantom{7}57
\nl		
Mrk\,1014 	&\phantom{XX1}8.7		&4.0	
	&\phantom{$<$}0.54	&\phantom{10}2.22	&\phantom{11}2.16
	&1.03		&2.14	&\phantom{1}53		&245
\nl
\nl
02483$+$4302 	&\phantom{XX1}2.9		&1.3 	
	&\phantom{$<$}0.19	&\phantom{10}4.02	&\phantom{11}6.92
	&0.58		&0.41	&\phantom{1}30		&142
\nl		
03158$+$4227 	&\phantom{XX1}7.0		&3.2 	
	&\phantom{$<$}0.45	&\phantom{10}4.26	&\phantom{11}4.28
	&1.00		&2.75	&\phantom{1}85		&392
\nl		
03521$+$0028 	&\phantom{XX1}9.3		&4.3 	
	&\phantom{$<$}0.23	&\phantom{10}2.64	&\phantom{11}3.84
	&0.69		&2.26	&\phantom{1}52		&242
\nl		
04232$+$1436 	&\phantom{XX1}9.0		&4.1 	
	&$<$0.38		&\phantom{10}3.45	&\phantom{11}4.26
	&0.81		&0.75	&\phantom{1}18		&\phantom{7}83
\nl		
VII\,Zw\,31 	&\phantom{XX}11.8			&5.4 	
	&\phantom{$<$}0.58	&\phantom{10}5.58	&\phantom{11}9.62
	&0.58		&0.64	&\phantom{1}11		&\phantom{7}53
\nl		
\nl		
07598$+$6508 	&\phantom{XX}11.4			&5.2 	
	&\phantom{$<$}0.53	&\phantom{10}1.69	&\phantom{11}1.73
	&0.98		&1.34	&\phantom{1}25		&118
\nl
08030$+$5243 	&\phantom{XX1}8.8		&4.1 	
	&\phantom{$<$}0.18	&\phantom{10}2.99	&\phantom{11}4.39
	&0.68		&0.75	&\phantom{1}18		&\phantom{7}85
\nl
08572$+$3915 	&\phantom{XX1}1.3		&0.6 	
	&\phantom{$<$}1.70	&\phantom{10}7.43	&\phantom{11}4.59
	&1.62		&1.00	&166			&766
\nl
09320$+$6134 	&\phantom{XX1}4.7		&2.1 	
	&\phantom{$<$}1.03	&\phantom{1}11.54	&\phantom{1}20.23
	&0.57		&0.69	&\phantom{1}32		&148
\nl		
10035$+$4852 	&\phantom{XX1}7.0		&3.2 	
	&\phantom{$<$}0.28	&\phantom{10}4.59	&\phantom{11}6.24
	&0.74		&0.67	&\phantom{1}20		&\phantom{7}94
\nl
\nl		
10190$+$1322 	&\phantom{XX1}8.2		&3.8 	
	&\phantom{$<$}0.38	&\phantom{10}3.35	&\phantom{11}5.57
	&0.60		&0.75	&\phantom{1}19		&\phantom{7}91
\nl		
10495$+$4424 	&\phantom{XX1}8.2		&3.8 	
	&\phantom{$<$}0.16	&\phantom{10}3.53	&\phantom{11}5.41
	&0.65		&1.12	&\phantom{1}29		&136
\nl		
10565$+$2448 	&\phantom{XX1}5.6		&2.6 	
	&\phantom{$<$}1.14	&\phantom{1}12.12	&\phantom{1}15.13
	&0.80		&0.76	&\phantom{1}29		&135
\nl		
11506$+$1331 	&\phantom{XX1}7.5		&3.4 	
	&\phantom{$<$}0.19	&\phantom{10}2.58	&\phantom{11}3.32
	&0.78		&1.47	&\phantom{1}42		&196
\nl		
Mrk\,231	&\phantom{XX1}7.5		&3.5 	
	&\phantom{$<$}8.66	&\phantom{1}31.99	&\phantom{1}30.29
	&1.06		&1.96	&\phantom{1}56		&259
\nl		
\nl		
13106$-$0922 	&\phantom{XX1}9.4		&4.3 	
	&$<$0.36	&\phantom{10}1.66	&\phantom{11}2.42
	&0.69		&1.89	&\phantom{1}43		&201
\nl		
Arp\,193 	&\phantom{XX1}3.8		&1.8 	
	&\phantom{$<$}1.36	&\phantom{1}15.44	&\phantom{1}25.18
	&0.61		&0.31	&\phantom{1}17		&\phantom{7}81
\nl		
Mrk\,273 	&\phantom{XX1}5.1		&2.3 	
	&\phantom{$<$}2.28	&\phantom{1}21.74	&\phantom{1}21.38
	&1.02		&1.06	&\phantom{1}43		&202
\nl		
13442$+$2321 	&\phantom{XX1}5.2		&2.4 	
	&\phantom{$<$}0.11	&\phantom{10}1.62	&\phantom{11}2.26
	&0.72		&1.18	&\phantom{1}49		&227
\nl
14070$+$0525	&\phantom{XX1}9.5		&4.4 	
	&\phantom{$<$}0.19	&\phantom{10}1.45	&\phantom{11}1.82
	&0.80		&3.80	&\phantom{1}87		&401		
\tablebreak
\nl
15030$+$4835 	&\phantom{XX}12.5\phantom{XX}	&5.8 	
	&$<$0.08		&\phantom{10}0.90	&\phantom{11}1.46
	&0.62	&1.70		&\phantom{1}29			&135 	
\nl		
Arp\,220 	&\phantom{XX1}7.0		&3.2 	
	&\phantom{$<$}7.91		&103.80		&112.40
	&0.92		&1.16	&\phantom{1}35		&164
\nl		
16090$-$0139	&\phantom{XX}12.2			&5.6 	
	&\phantom{$<$}0.26	&\phantom{10}3.61	&\phantom{11}4.87
	&0.74		&2.29	&\phantom{1}40		&188
\nl
16334$+$4630	&\phantom{XX1}9.5		&4.4 	
	&$<$0.10	&\phantom{10}1.19	&\phantom{11}2.09
	&0.57		&1.80	&\phantom{1}41		&189
\nl		
NGC\,6240 	&\phantom{XX1}7.9		&3.7 	
	&\phantom{$<$}3.42	&\phantom{1}22.68	&\phantom{1}27.78
	&0.82		&0.45	&\phantom{1}12		&\phantom{7}56
\nl
\nl		
17208$-$0014 	&\phantom{XX1}7.1		&3.2 	
	&\phantom{$<$}1.66	&\phantom{1}34.14	&\phantom{1}34.90
	&0.98		&2.14	&\phantom{1}65		&302
\nl
18368$+$3549 	&\phantom{XX1}8.5		&3.9 	
	&$<$0.25		&\phantom{10}2.23	&\phantom{11}3.84
	&0.58		&1.20	&\phantom{1}30		&140
\nl		
19297$-$0406	&\phantom{XX1}9.4		&4.3 	
	&\phantom{$<$}0.59	&\phantom{10}7.05	&\phantom{11}7.72
	&0.91		&1.80	&\phantom{1}41		&191
\nl		
19458$+$0944	&\phantom{XX}12.0		&5.5 	
	&$<$0.28		&\phantom{10}3.94	&\phantom{11}7.11
	&0.55		&1.59	&\phantom{1}28		&132
\nl		
20087$-$0308 	&\phantom{XX}16.1			&7.4 	
	&\phantom{$<$}0.24	&\phantom{10}4.70	&\phantom{11}6.54
	&0.72		&1.87	&\phantom{1}25		&116
\nl
\nl		
22542$+$0833 	&\phantom{XX1}6.0		&2.8 	
	&$<$0.18	&\phantom{10}1.20	&\phantom{11}1.48
	&0.81		&1.18	&\phantom{1}42		&196
\nl		
23365$+$3604 	&\phantom{XX1}8.5		&3.9 	
	&\phantom{$<$}0.81	&\phantom{10}7.09	&\phantom{11}8.36
	&0.85		&1.01	&\phantom{1}25		&119
\nl
\nl
{\it median:}	&\phantom{XX1}8\phantom{.0}		& --- 	
	&\phantom{$<$} ---	&\phantom{10} ---	&\phantom{11} ---
	&0.8\phantom{5}		&1.3\phantom{1}	
	&\phantom{1}35		&160
%
\tablenotetext{}{
$L_{\rm FIR} =
3.94\times 10^5\, r(S_{60}/S_{100})\cdot (2.58\, S_{60} + S_{100})\, D_L^2$, 
with  $L_{\rm FIR}$ in \Lsun , $S$ in Jy,  $D_L$ in Mpc,
}
\tablenotetext{}{
where $D_L = c H_0^{-1} q_0^{-2} \left\{ z q_0 + \left( q_0 - 1 \right)
	\left[ \left( 2 q_0 z + 1 \right)^{0.5} - 1 \right] \right\}$
	\ \ \  (e.g., Weinberg 1972).
}
\tablenotetext{}{
We adopt $H_0$=75 km\ts s$^{-1}\ts$Mpc$^{-1}$ and $q_0 = 0.5$ .
}
\tablenotetext{}{
L$_l \equiv $ \Kkmspc , \ \ \ 
$M^\prime({\rm H}_2)= 4.6 \cdot L^\prime_{\rm CO}$ .
}
\end{planotable}
\clearpage

\begin{planotable}{lllllll}
\tablecaption{Radii and Dynamical Masses Derived from CO Data}
\label{Radii and Dynamical Masses Derived from CO Data}
\tablehead{
\colhead{Source}&\colhead{bb Temp.}		&\colhead{bb radius}
		&\colhead{CO radius}		&\colhead{Dyn.\ Mass}	
	&\colhead{$	\displaystyle \strut {  M_{\rm dyn} \over 
				\displaystyle \strut L^\prime_{\rm CO}  }   $}
		&\colhead{Density}
\nl
\colhead{name}	&\colhead{$T_{\rm bb}$}	&\colhead{$R_{\rm bb}$} 
	&\colhead{$R_{\rm CO}$}	&\colhead{ $M_{\rm dyn}$($<R_{\rm CO}$)}
	        &\colhead{}
                &\colhead{$n_{\rm tot}$} 
\nl
\colhead{}	&\colhead{(K)}			&\colhead{(pc)}	
		&\colhead{(pc)}			&\colhead{(10$^8$\,\Msun )}
		&\colhead{$ 	
			\displaystyle \left(\strut { {\rm M}_\odot 
		\over	\displaystyle \strut   {\rm L}_l  } \right)    
			  $}
		&\colhead{(cm$^{-3}$)}
}
\startdata
00057$+$4021 	&\phantom{1}72		&\phantom{1}81	
		&219			&\phantom{1}62		
		&\phantom{11}1.6        &2850
\nl		
00188$-$0856 	&\phantom{1}62		&238	
		&297			&105		
		&\phantom{11}1.6        &1932
\nl		
00262$+$4251	&\phantom{1}88		&\phantom{1}98	
		&314			&\phantom{1}66		
		&\phantom{11}1.1       	&1023
\nl
I Zw 1  	&\phantom{1}63		&101	
		&248			&\phantom{1}97		
		&\phantom{11}1.9        &3060
\nl		
Mrk\,1014 	&\phantom{1}80		&174	
		&418			&\phantom{1}87		
		&\phantom{11}1.0        &\phantom{1}576
\nl
\nl
02483$+$4302 	&\phantom{1}50		&197	
		&273			&\phantom{1}57		
		&\phantom{11}2.0        &1352
\nl		
03158$+$4227 	&\phantom{1}75		&220	
		&405			&\phantom{1}85		
		&\phantom{11}1.2        &\phantom{1}613
\nl		
03521$+$0028 	&\phantom{1}60		&319	
		&575			&120		
		&\phantom{11}1.3        &\phantom{1}304
\nl		
04232$+$1436 	&\phantom{1}62		&172	
		&340			&126		
		&\phantom{11}1.4        &1544
\nl		
VII\,Zw\,31 	&\phantom{1}50		&244	
		&616			&129		
		&\phantom{11}1.1        &\phantom{1}265
\nl		
\nl		
07598$+$6508 	&\phantom{1}76		&153	
		&377			&\phantom{1}99		
		&\phantom{11}0.9        &\phantom{1}893
\nl
08030$+$5243 	&\phantom{1}56		&208	
		&345			&141		
		&\phantom{11}1.6        &1653
\nl
08572$+$3915 	&119			&\phantom{1}53	
		&113			&\phantom{1}24		
		&\phantom{11}1.8        &7862
\nl
09320$+$6134 	&\phantom{1}49		&266	
		&295			&\phantom{1}84		
		&\phantom{11}1.8        &1570
\nl		
10035$+$4852 	&\phantom{1}58		&186	
		&395			&\phantom{1}82		
		&\phantom{11}1.2        &\phantom{1}646
\nl
\nl		
10190$+$1322 	&\phantom{1}52		&243	
		&360			&127		
		&\phantom{11}1.5        &1316
\nl		
10495$+$4424 	&\phantom{1}55		&265	
		&379			&\phantom{1}96		
		&\phantom{11}1.2        &\phantom{1}846
\nl		
10565$+$2448 	&\phantom{1}59		&188	
		&317			&\phantom{1}66		
		&\phantom{11}1.2        &1000
\nl		
11506$+$1331 	&\phantom{1}63		&229	
		&361			&\phantom{1}75		
		&\phantom{11}1.0        &\phantom{1}772
\nl		
Mrk\,231	&\phantom{1}73		&198	
		&378			&\phantom{1}79		
		&\phantom{11}1.0        &\phantom{1}702
\nl		
\nl		
13106$-$0922 	&\phantom{1}61		&281	
		&496			&104		
		&\phantom{11}1.1        &\phantom{1}410
\nl		
Arp\,193 	&\phantom{1}50		&170	
		&244			&\phantom{1}95		
		&\phantom{11}2.5        &3156
\nl		
Mrk\,273 	&\phantom{1}70		&157	
		&280			&\phantom{1}59		
		&\phantom{11}1.2        &1284
\nl		
13442$+$2321 	&\phantom{1}61		&221	
		&441			&\phantom{1}92		
		&\phantom{11}1.8        &\phantom{1}518
\nl
14070$+$0525	&\phantom{1}72		&284	
		&394			&\phantom{1}82		
		&\phantom{11}0.9        &\phantom{1}648		
\tablebreak
\nl
15030$+$4835 	&\phantom{1}59		&281	
		&499			&104		
		&\phantom{11}0.8	&\phantom{1}404
\nl		
Arp\,220 	&\phantom{1}64		&197	
		&270			&144		
		&\phantom{11}2.1	&3549
\nl		
16090$-$0139	&\phantom{1}61		&304	
		&459			&\phantom{1}96		
		&\phantom{11}0.8        &\phantom{1}477
\nl
16334$+$4630	&\phantom{1}56		&327	
		&413			&\phantom{1}98		
		&\phantom{11}1.0        &\phantom{1}673
\nl		
NGC\,6240 	&\phantom{1}59		&146	
		&341			&108		
		&\phantom{11}1.4        &1320
\nl
\nl		
17208$-$0014 	&\phantom{1}69		&235	
		&301			&\phantom{1}91		
		&\phantom{11}1.3        &1602
\nl
18368$+$3549 	&\phantom{1}53		&299	
		&396			&100		
		&\phantom{11}1.2        &\phantom{1}775
\nl		
19297$-$0406	&\phantom{1}68		&222	
		&385			&\phantom{1}80		
		&\phantom{11}0.9        &\phantom{1}680
\nl		
19458$+$0944	&\phantom{1}51		&373	
		&465			&132		
		&\phantom{11}1.1        &\phantom{1}633
\nl		
20087$-$0308 	&\phantom{1}59		&297	
		&466			&173		
		&\phantom{11}1.1        &\phantom{1}824
\nl
\nl		
22542$+$0833 	&\phantom{1}67		&182	
		&436			&\phantom{1}91		
		&\phantom{11}1.5        &\phantom{1}529
\nl		
23365$+$3604 	&\phantom{1}63		&191	
		&372			&\phantom{1}83		
		&\phantom{11}1.0        &\phantom{1}777
%
\tablenotetext{}{
$T_{\rm bb} \approx 
-(1+z) \displaystyle \strut \left( 
{ 82 \over \displaystyle \strut {\ln (0.3\cdot S_{60}/S_{100} )} } -0.5  
\right); \ \ \ \ \   \ \  
R_{\rm bb} 
	= (L_{\rm FIR}/(4\pi \sigma T_{\rm bb}^4)^{0.5}\ \ .
$}
\tablenotetext{}{
$R_{\rm CO}({\rm min}) 
	= (L^\prime_{\rm CO}/(\pi T_{\rm bb} \Delta V)^{0.5}\ \ ; 
\ \ \ \ \   \ \ \ \ \   \ \ \ \ \  
M_{\rm dyn}(<R_{\rm CO}) = 232\cdot R_{\rm CO}\cdot 
[{\rm Max}(300, \Delta V)]^2 
\ \ .$
}
\end{planotable}
\clearpage

\begin{planotable}{lrrrr}
\tablecaption{Mass Estimates for Ultraluminous Galaxies}
\label{Mass Estimates for Ultraluminous Galaxies}
\tablehead{
\colhead{Source}&\colhead{ $M_{\rm dyn}$($<R_{\rm CO}$)}&
\colhead{$M^\prime({\rm H}_2)$}&\colhead{$M_{\rm thin}$}&
\colhead{$100 M_{\rm dust}$}
\nl
\colhead{name}&\colhead{(10$^8$\,\Msun )}&\colhead{(10$^8$\,\Msun )}
&\colhead{(10$^8$\,\Msun )}&\colhead{(10$^8$\,\Msun )}
}
\startdata
  00057$+$4021 &         62 &        175 &         19 &      15 \nl
  00188$-$0856 &        105 &        309 &         29 &     130 \nl
  00262$+$4251 &         65 &        286 &         38 &      28 \nl
  I\,Zw\,1     &         97 &        230 &         22 &      22 \nl
  Mrk\,1014    &         87 &        402 &         48 &      76 \nl
 \nl
  02483$+$4302 &         57 &        133 &         10 &     82 \nl
  03158$+$4227 &         85 &        322 &         37 &    115 \nl
  03521$+$0028 &        120 &        427 &         38 &    207 \nl
  04232$+$1436 &        126 &        413 &         38 &     61 \nl
  VII\,Zw\,31  &        129 &        544 &         41 &    111 \nl
 \nl
  07598$+$6508 &         99 &        523 &         60 &     57 \nl
  08030$+$5243 &        141 &        406 &         34 &     44 \nl
  08572$+$3915 &         24 &         60 &         11 &     46 \nl
  09320$+$6134 &         84 &        214 &         16 &     26 \nl
  10035$+$4852 &         82 &        324 &         28 &    119 \nl
 \nl
  10190$+$1322 &        127 &        378 &         30 &     88 \nl
  10495$+$4424 &         96 &        376 &         31 &    168 \nl
  10565$+$2448 &         66 &        258 &         23 &     82 \nl
  11506$+$1331 &         75 &        345 &         33 &    148 \nl
  Mrk\,231     &         79 &        347 &         38 &     92 \nl
 \nl
  13106$-$0922 &        103 &        431 &         40 &    160 \nl
  Arp\,193     &         95 &        176 &         13 &     57 \nl
  Mrk\,273     &         59 &        230 &         25&     56 \nl
  13442$+$2321 &         92 &        239 &         22 &    106 \nl
  14070$+$0525 &         82 &        435 &         47 &    188 \nl
\tablebreak
\nl
  15030$+$4835 &        104 &        575 &         51 &    164 \nl
  Arp\,220     &        144 &        323 &         31 &     84 \nl
  16090$-$0139 &         96 &        560 &         52 &    195 \nl
  16334$+$4630 &         98 &        438 &         37 &    208 \nl
  NGC\,6240    &        108 &        366 &         33 &     44 \nl
 \nl
  17208$-$0014 &         91 &        325&         34 &    123 \nl 
  18368$+$3549 &        100 &        393 &         31 &    171 \nl
  19297$-$0406 &         80 &        434 &         44 &    107 \nl
  19458$+$0944 &        132 &        552 &         42 &    261 \nl
  20087$-$0308 &        173 &        740 &         66 &    184 \nl
 \nl
  22542$+$0833 &         91 &        277 &         28 &    101 \nl
  23365$+$3604 &         83 &        390 &         37 &     77 \nl
\end{planotable}
\clearpage

\begin{planotable}{lcclccl}
\tablecaption{Optical / Near-IR Morphology of Ultraluminous Galaxies}
\label{Optical / Near-IR Morphology of Ultraluminous Galaxies}
\tablehead{
\colhead{Source}&\colhead{Redshift\phantom{XX}} &\colhead{Distance} 
		&\colhead{Morphology} 		&\colhead{Separation}	
		&\colhead{}		&\colhead{Ref.}
\nl
\colhead{name}	&\colhead{$cz$}			&\colhead{$D_A$}	
		&\colhead{}			&\colhead{$\theta$}
                &\colhead{$D_A\cdot \theta$}  	&\colhead{}
\nl
\colhead{}	&\colhead{(\,km\, s$^{-1}$)}	&\colhead{(Mpc)}	
		&\colhead{}			&\colhead{(arcsec)}
		&\colhead{(kpc)}		&\colhead{}
}
\startdata
00057$+$4021 	&\phantom{6}13390		&165		
		&Double? disturbed		&\phantom{~?}2.5~?
		&\phantom{~?}2\phantom{.5}~?
          	&AHM87
\nl		
00188$-$0856 	&\phantom{6}38530		&415		
		&Merger $+$ companion		&7\phantom{.5}
		&14\phantom{1.5}
        	&M96
		
\nl		
00262$+$4251	&\phantom{6}29153		&330		
		&Single  			&$<$0.8\phantom{$<$}
		&$<$1.3\phantom{$<$}
		&M96
		
\nl
I Zw 1  	&\phantom{6}18330		&220		
		&Single (quasar)		&---
		&---				&
		
\nl		
Mrk\,1014 	&\phantom{6}48947		&501		
		&Single (quasar)		&---
		&---				&
		
\nl
\nl
02483$+$4302 	&\phantom{6}15419		&188		
		&Merger				&3.8
		&3.5	
		&KDBS
		
\nl		
03158$+$4227 	&\phantom{6}40296		&431		
		&Single 			&$<$0.8\phantom{$<$}
		&$<$1.6\phantom{$<$}	
		&M96
		
\nl		
03521$+$0028 	&\phantom{6}45530		&474		
		&Double 			&1.6
		&3.6     		        &M96
		
\nl		
04232$+$1436 	&\phantom{6}23855		&278		
		&Double, merger			&4.6
		&5\phantom{.5}
		&
		
\nl		
VII\,Zw\,31 	&\phantom{6}16260		&196		
		&Single (starburst)		&---
		&---				&
		
\nl		
\nl		
07598$+$6508 	&\phantom{6}44621		&466		
		&Single (quasar)		&---
		&---				&HB89, S92
		
\nl
08030$+$5243 	&\phantom{6}25031		&290		
		&Single				&$<$0.8\phantom{$<$}
		&$<$1.1\phantom{$<$}		&M96
		
\nl
08572$+$3915 	&\phantom{6}17450		&211		
		&Double				&6\phantom{.5}
		&5.5	         		&S92
		
\nl
09320$+$6134 	&\phantom{6}11785		&147		
		&Single; tails			&$<$1.0\phantom{$<$}
		&$<$0.7\phantom{$<$}		&S92, M96
		
\nl		
10035$+$4852 	&\phantom{6}19427		&232		
		&---				&---
		&---				&
		
\nl
\nl		
10190$+$1322 	&\phantom{6}22953		&269		
		&Double				&5\phantom{.5}
		&6.6            		&Fig.\ 7
		
\nl		
10495$+$4424 	&\phantom{6}27674		&315		
		&Single, distorted		&$<$0.7\phantom{$<$}
		&$<$1.1\phantom{$<$}		&Fig.\  7, M96
		
\nl		
10565$+$2448 	&\phantom{6}12923		&160		
		&Double, Multiple		&\phantom{26}8,\phantom{5}26
		&\phantom{20}6,\phantom{5}20    &Fig.\ 7, M96
		
\nl		
\hbox{11506$+$1331} 	&\phantom{6}38158		&412		
		&Double, Single			&$<$0.5\phantom{$<$}
		&$<$1.1\phantom{$<$} 		&Fig.\  7, M96
		
\nl		
Mrk\,231	&\phantom{6}12650		&157		
		&Double nucl.?, tail		&3.5
		&2.8				&A94
		
\nl		
\nl		
13106$-$0922 	&\phantom{6}52290		&526		
		&Double				&2\phantom{.5}
		&5\phantom{.5}			&Fig.\  7
		
\nl		
Arp\,193 	&\phantom{60}7000		&\phantom{1}90		
		&Single $+$ 2 tails             &\phantom{~?}$>$1\phantom{.5}~?\phantom{$<$}
		&\phantom{~?}$>$4\phantom{.5}~?\phantom{$>$}	
        	&S92
		
\nl		
Mrk\,273 	&\phantom{6}11324		&142		
		&Double in NIR, 2 tails		&0.9
		&0.6				&M93
		
\nl		
13442$+$2321 	&\phantom{6}42620		&450		
		&Merger?			&---
		&---				&
		
\nl
14070$+$0525	&\phantom{6}79621		&702		
		&Single				&---
		&---				&Fig.\  7	
\tablebreak
\nl
15030$+$4835 	&\phantom{6}64900		&613		
		&Double				&3.5
		&11\phantom{.5}			&Fig.\  7
		
\nl		
Arp\,220 	&\phantom{60}5450		&\phantom{1}70		
		&Merger, double nuc.?		&1\phantom{.5}
		&0.3				&G90, M93, S94
		
\nl		
16090$-$0139	&\phantom{6}40044		&429		
		&Single, distorted		&$<$1.0\phantom{$<$}
		&$<$2.2\phantom{$<$}		&Fig.\  7, M96
		
\nl
16334$+$4630	&\phantom{6}57250		&562		
		&Double				&4.4
		&12.6\phantom{1}		&Fig.\  7
		
\nl		
NGC\,6240 	&\phantom{60}7298	        &\phantom{1}93		
		&Double nucl.			&---
		&---				&AHM90; L94	
		
\nl
\nl		
17208$-$0014 	&\phantom{6}12837		&159		
		&Double +tails, Single 		&$<$0.8\phantom{$<$}
		&$<$0.6\phantom{$<$}      	&Fig.\  7, M96

\nl
18368$+$3549 	&\phantom{6}34850		&383		
		&Single				&$<$0.8\phantom{$<$}
		&$<$1.5\phantom{$<$}		&M96

\nl		
19297$-$0406	&\phantom{6}25700		&297		
		&Single				&$<$1.9\phantom{$<$}
		&$<$2.7\phantom{$<$}		&Fig.\  7, M96
		
\nl		
19458$+$0944	&\phantom{6}29980		&338		
		&Double?, Single		&$<$0.8\phantom{$<$}
		&$<$1.4\phantom{$<$}		&M96
		
\nl		
20087$-$0308 	&\phantom{6}31688		&354		
		&Single, tails			&$<$1.0\phantom{$<$}
		&$<$1.7\phantom{$<$}		&MM90, M96
		
\nl
\nl		
22542$+$0833 	&\phantom{6}49750		&487		
		&Double				&---
		&---				&
		
\nl		
23365$+$3604 	&\phantom{6}19330		&231		
		&Single, tails			&$<$0.9\phantom{$<$}
		&$<$1.1\phantom{$<$}		&Fig.\  7, M96
%
\tablenotetext{}{
$D_A = D_L / (1+z)^2$; for $D_L$, see footnote to Table~2. 
}
\tablenotetext{}{
A87, A90, A94: Armus et al.\  (1987; 1990; 1994);  G90: Graham et al.\  (1990);
}
\tablenotetext{}{
KDBS: Kollatschny et al.\  (1991); L94: Lester \& Gaffney\ (1994); 
}
\tablenotetext{}{
M93: Majewski et al.\  (1993); M96: Murphy et al.\  (1996);
}
\tablenotetext{}{
MM90: Melnick \& Mirabel\  (1990); S92: Sanders\  (1992);
}
\tablenotetext{}{
S94: Shaya et al.\  (1994).
}
\end{planotable}
\clearpage



\clearpage

\newpage
\begin{figure}
\caption[Redshift histogram]
{
Distribution of redshifts in our sample of ultraluminous galaxies.
}
\end{figure}
\begin{figure}
\caption[CO(1--0) Spectra]
{
CO(1--0) spectra of galaxies in our sample.  The vertical scale is main-beam 
brightness temperature, in mK.  Source names and redshifts are indicated in
the boxes.  For each source, the zero of the (Doppler) velocity scale is
relative to the redshifts listed in Table~1.  The spectra are in order of
decreasing redshift.
}
\end{figure}
\begin{figure} 
\caption[\Lfir\ vs. \Lco ]
{
FIR luminosity vs. CO(1--0) luminosity (lower scale) and molecular gas mass
(top scale). Solid circles indicate galaxies in our sample.  Normal and weakly
interacting spirals are scattered about the solid line.  Open circles show 
locations in this diagram for some well-known galaxies. 
}
\end{figure}
\begin{figure}
\caption[\Ico\ vs. 100\,$\mu$m flux density] 
{
Integrated CO(1--0) line intensity, \Ico , in K$_{\rm mb}$\,\kms , vs. 
IRAS 100\,$\mu$m flux density, in Jy, for the ultraluminous galaxies in our
sample.  The solid line corresponds to the relation 
\Ico\ $= 1.0 \times S_{100}$.
}
\end{figure}
\begin{figure}
\caption[$L_{\rm FIR}/L_{\rm CO}$ vs. $T_bb$]
{
The distant-independent quantity $L_{\rm FIR}/L_{\rm CO}$ vs. the 
far~IR black body dust temperature derived from the IRAS flux
ratio $S_{60}/S_{100}$. 
The line indicates the upper limit for the luminosity ratio for the 
black body model in eq.(11), with $f_V\,\Delta V = 300$\,\kms .
Solid circles: galaxies in our sample with $L_{\rm FIR} > 10^{12}$\,\Lsun .  
Open circles: galaxies with $L_{\rm FIR} < 10^{12}$\,\Lsun .
}
\end{figure}
\begin{figure}
\caption[histogram of $\Delta V$'s] 
{
Distribution of observed CO full widths to half-maximum, 
$\Delta V$, for the galaxies in our sample.
}
\end{figure}
\begin{figure}
\caption[R-band CCD images]
{
R-band CCD images of galaxies between RA 10h and 23h 
in the sample.  The images are in  order of decreasing redshift.
 North is at the top, east is to
the left.  Brightness contours are on an arbitrary linear scale.  Next to each
image is source name, redshift and CO(1--0) spectrum, with units as in Fig.~1.
The images were taken at the Univ. of Hawaii 88-in. telescope by Sanders \&
Kim (1993, priv. communication).  
}
\end{figure}
\clearpage

\begin{references}
\reference Armus, L., Heckman, T.M., \& Miley, G.K. 1987, AJ, 94, 831

\reference Armus, L., Heckman, T.M., \& Miley, G.K.  1990, ApJ, 364, 471

\reference Armus, L., Surace, J.A., Soifer, B.T., Matthews, K., 
Graham, J.R., \& Larkin, J.E. 1994, AJ, 108, 76

\reference Barvainis, R., Alloin, D., \& Antonucci, R. 1989, ApJ, 337, L69 

\reference Beichman, C.A., Neugebauer, G., Habing, H.J., Clegg, P.E., 
Chester, T.J. (eds.) 1988, IRAS Catalogs and Atlases Explanatory Supplement,
(Washington, D.C.: U.S.\ Govt.\ Printing Office)

\reference Bryant, P.M., \& Scoville, N.Z. 1996, ApJ, 457, 692

\reference Condon, J.J., Huang, Z.P., Yin, Q.F., \& Thuan, T.X.  1991, ApJ,
378, 65

\reference Downes, D., Solomon, P.M., \& Radford, S.J.E. 1993, ApJ, 414, L13
(DSR).

\reference Fisher, K.B., Huchra, J.P., Strauss, M.A., Davis, M., Yahil, A.,
\& Schlegel, D. 1995, ApJS, 100, 69


\reference Graham, J.R., Carico, D.P., Matthews, K., Neugebauer, G., 
Soifer, B.T., \& Wilson, T.D. 1990, ApJ, 354, L5

\reference Hildebrand, R.H. 1983, QJRAS, 24, 267

\reference Infrared Astronomical Satellite (IRAS) Point Source Catalog, 
Version 2  1988, Joint IRAS Science Working Group, 
(Washington, D.C.: U.S.\ Govt.\ Printing Office)

\reference Jog, C.J., \&  Solomon, P.M. 1992, ApJ, 387, 152

\reference Jog, C.J., \& Das, M. 1992, ApJ, 400, 476

\reference Kollatschny, W., Dietrich, M., Borgeest, U., \& 
Schramm, K.J. 1991, A\&A, 249, 57

\reference Lester, D.F., \& Gaffney, N.I. 1994, ApJ, 431, L13 

\reference Lonsdale, C.J., Persson, S.E., \& Matthews, K. 1984, ApJ, 287, 95

\reference Lonsdale, C.J., Helou, G., Good, J.C., \& Rice, W. 1985,
in Cataloged Galaxies and Quasars Observed in the IRAS Survey, 
(Pasadena: JPL)

\reference Majewski, S.R., Hereld, M., Koo, D.C., Illingworth, G.D., 
\& Heckman, T.M. 1993, ApJ, 402, 76 

\reference Melnick, J.\ \& Mirabel, I.F. 1990, A\&A, 231, L19

\reference Mihos, J.C., \& Hernquist, L. 1994, ApJ, 431, L9

\reference Mooney, T. \& Solomon, P.M. 1988, ApJ, 334, L51

\reference Moshir et al.\  1992, IRAS Faint Source Survey Explanatory Supplement,
Version 2, (Pasadena: JPL)

\reference Murphy, T.W.~Jr., Armus L., Matthews, K., Soifer, B.T.,
 Mazzarella, J.M., \& Shupe, D.L.  1996, AJ, 111, 1025

\reference Norman,. C. A. 1991, in Massive Stars in Starbursts, 
ed. C.\ Leitherer, N.R.\ Walborn, T.M.\ Heckman, \& C.A.\ Norman, 
(Cambridge: Camb. Univ. Press) 271

\reference Okumura, S.K., Kawabe, R., Ishiguro, M., Kasuga, T., Morita, K.I.,
\& Ishizuki, S.  1991, in Dynamics of Galaxies and Their Molecular Cloud
Distributions, ed.\ F.\ Combes \& F.\ Casoli (Dordrecht: Kluwer), 425

\reference Planesas, P., Mirabel, I.F.,  \& Sanders, D.B.  1991, ApJ, 370, 172

\reference Radford, S.J.E., Solomon, P.M., \& Downes, D. 1991a, ApJ, 368, L15

\reference Radford, S.J.E., et al.  1991b,  in Dynamics of Galaxies and Their 
Molecular Cloud Distributions, ed.\ F.\ Combes \& F.\ Casoli 
(Dordrecht: Kluwer), 303

\reference Sanders, D.B. 1992, in Relationships Between Active Galactic Nuclei
and Starburst Galaxies, ASP Conf.\ Ser.\ 31, ed. A.V.\ Filippenko, 
(San Francisco: ASP) 303

\reference Sanders, D.B., Soifer, B.T., Elias, J.H., Madore, B.F., 
Matthews, K., Neugebauer, G., \& Scoville, N.Z.  1988a, ApJ, 325, 74

\reference Sanders, D.B., Soifer, B.T., Elias, J.H., Neugebauer, G.,
\& Matthews, K. 1988b, ApJ, 328, L35

\reference Sanders, D.B., Scoville, N.Z., \& Soifer, B.T. 1991, ApJ, 370, 158

\reference Sanders, D.B., et al., 1986, ApJ, 305, L45

\reference Scoville, N.Z., Sanders, D.B., \& Clemens, D.P., 1986,
ApJ, 310, L77

\reference Scoville, N.Z., Sargent, A.I., Sanders, D.B., \& Soifer, B.T. 1991,
ApJ, 366, L5

\reference Shaya, E.J., Dowling, D.M., Currie, D.G., Faber, S.M., \& 
Groth, E.J.
1994, AJ, 107, 1675

\reference Shier, L.M., Rieke, M.J., \& Rieke, G.H. 1994, ApJ, 433, L9

\reference Soifer, B.T., et al.\  1984, ApJ, 278, L71

\reference Solomon, P.M., Downes, D., \& Radford, S.J.E. 1992a, ApJ, 387, L55

\reference Solomon, P.M., Radford, S.J.E., \& Downes, D. 1990, ApJ, 348, L53

\reference Solomon, P.M., Radford, S.J.E., \& Downes, D. 1992, Nature, 356,
318
\reference Solomon, P.M., \& Rivolo, A.R. 1989, ApJ, 339, 919

\reference Solomon, P.M., Rivolo, A.R., Barrett, J.W., \& Yahil, A.
1987, ApJ, 319, 730

\reference Solomon, P.M., \& Sage, L.J. 1988, ApJ, 334, 613

\reference Strauss, M.A., Huchra, J.P., Davis, M.,  Yahil, A., 
Fisher, K.B., \& Tonry, J. 1992, ApJS, 83, 29

\reference Wright, G.S., Joseph, R. D. \& Meikle, W.P. 1984, Nature, {\bf 344} 417

\reference  Weinberg, S. 1972, Gravitation and Cosmology (New York: Wiley)
 
\end{references}
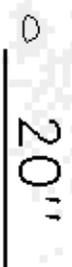
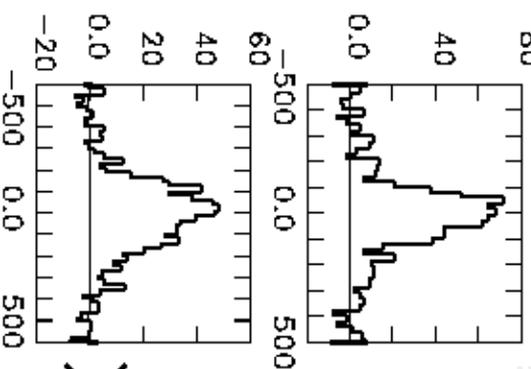
\end{document}